\documentclass[11pt,letterpaper]{article}

\usepackage[T1]{fontenc}
\usepackage[margin=1in]{geometry}
\usepackage{microtype}

\usepackage{amsmath,amsfonts,amssymb,amsthm,mathrsfs,mathtools}
\usepackage{libertinus}
\usepackage{libertinust1math}
\usepackage{dsfont}
\usepackage[thinc]{esdiff}
\numberwithin{equation}{section}

\usepackage{booktabs}
\usepackage[ruled,noend]{algorithm2e}
\usepackage{tikz}
\usetikzlibrary{arrows.meta,positioning,calc,fit,backgrounds}
\usepackage{xcolor}
\usepackage[most]{tcolorbox}
\usepackage{multicol}
\usepackage{tocloft}
\usepackage{titlesec}

\usepackage[round]{natbib}
\usepackage{color-edits}
\addauthor{ss}{magenta}

\definecolor{SkyBlueDark}{HTML}{246F9A}    
\definecolor{SkyBlueAccent}{HTML}{61ADD8}  
\definecolor{SkyBlueText}{HTML}{536F80}    
\definecolor{SkyBlueDots}{HTML}{C6DCE9}    
\definecolor{SkyBlueRule}{HTML}{D6EAF5}    

\usepackage[hyperindex,breaklinks]{hyperref}
\hypersetup{
  colorlinks=true,
  citecolor=SkyBlueDark,
  linkcolor=SkyBlueDark,
  urlcolor=SkyBlueDark,
  linktoc=page
}
\usepackage{packages/macros-common}
\usepackage{packages/macros-project-specific}
\usepackage{packages/mytheorems}
\definecolor{dividerSky}{RGB}{78, 163, 207}
\definecolor{dividerLight}{RGB}{211, 229, 239}

\definecolor{dividerSky}{RGB}{52, 157, 218}

\newcommand{\abstracttocdivider}{%
  \par\vspace{1.2em}
  \noindent
  \begin{tikzpicture}
    \shade[
      left color=dividerSky,
      right color=dividerSky!5!white
    ] (0,0) rectangle (\linewidth, 1pt);
  \end{tikzpicture}
  \par\vspace{1em}
}
\titlespacing*{\section}{0pt}{2.4ex plus 0.6ex minus 0.3ex}{1.0ex}
\titlespacing*{\subsection}{0pt}{1.8ex plus 0.5ex minus 0.2ex}{0.7ex}

\SetAlFnt{\small}
\SetAlCapFnt{\small}
\SetAlCapNameFnt{\small}
\SetAlCapHSkip{0pt}
\IncMargin{-\parindent}

\newcounter{algorithm}
\newenvironment{pseudocode}[2]{%
  \begin{figure}[htbp]\refstepcounter{algorithm}\label{#2}%
  \centering\begin{minipage}{0.96\linewidth}%
  \hrule\vspace{6pt}\textbf{Algorithm \thealgorithm. #1}\par\smallskip
  \small
}{\vspace{5pt}\hrule\end{minipage}\end{figure}}

\cftsetindents{section}{0em}{1.45em}
\cftsetindents{subsection}{1.45em}{2.10em}
\makeatletter
\newcommand{\SkyBlueTOC}{%
  \par\vspace{0pt}%
  \begingroup
    \fontsize{8.5}{11.2}\selectfont
    \setlength{\parskip}{0pt}%
    \setlength{\columnsep}{18pt}%
    \setlength{\columnseprule}{0.35pt}%
    \renewcommand{\columnseprulecolor}{\color{SkyBlueRule}}%
    \hyphenpenalty=10000 \exhyphenpenalty=10000
    \emergencystretch=1.2em
    \raggedright
    \setlength{\multicolsep}{0pt}%
    \begin{multicols}{2}
      \@starttoc{toc}%
    \end{multicols}%
  \endgroup
  \par
}
\makeatother

\title{Settling the Time and Structural Complexity of Bertrand Networks}
\newcommand{\authorblock}[3]{%
  \begin{tabular}[t]{@{}c@{}}
    {\fontsize{11.5}{14}\selectfont #1}\\[3pt]
    {\fontsize{8.5}{11}\selectfont\color{SkyBlueText} #2}\\[1.5pt]
    {\fontsize{8.5}{11}\selectfont\href{mailto:#3}{\nolinkurl{#3}}}
  \end{tabular}%
}
\author{%
  \authorblock{MohammadTaghi Hajiaghayi}
    {University of Maryland, College Park, USA}{hajiaghayi@gmail.com}%
  \hspace{2em}%
  \authorblock{Suho Shin}
    {Stanford University, Stanford, USA}{suhoyah@gmail.com}%
}

\begin{document}
\date{}
\maketitle
\begin{abstract}
In the seminal Bertrand network game introduced by~\cite{BLN13}, each seller posts a single price to captive buyers and to the markets it shares with neighboring sellers, where sellers cannot price discriminate across the markets.
Although the existence of equilibria and polynomial-time algorithms to compute them in trees~\citep{BLN13} have been known, it has been a major open problem whether there exist efficient algorithms to compute them for general graphs.

In this paper, we resolve this open question negatively by proving that computing any constant-error approximate Nash equilibrium in a Bertrand network is $\PPAD$-complete. This hardness holds even on sparse bipartite graphs of maximum degree five with constant integer market weights. Complementing our hardness result, we design an efficient polynomial-time approximation scheme (EPTAS) when both the maximum degree and the number of sellers with captive buyers are bounded.

Beyond time complexity, since exact equilibria require randomizing prices over continuous intervals, another fundamental question left open by~\cite{BLN13} is whether an exact equilibrium can always be represented with a finite number of price intervals, \ie whether its structural complexity is finite. 
We resolve this question affirmatively by proving that every achievable equilibrium revenue profile can be realized by an equilibrium with structural complexity $O(2^n)$, which leads to an exact algorithm running in doubly exponential time.
We further show that this bound is tight: there exist instances for which the unique equilibrium realizing target revenue profiles has structural complexity $\Omega(2^n)$.

\end{abstract}

\abstracttocdivider

\SkyBlueTOC





\clearpage
\section{Introduction}

Sellers in markets often face distinct competitors across different market segments. 
Accordingly, buyers in such shared markets compare prices across different sellers.
Further, in many scenarios, sellers cannot discriminate prices over all market segments, including not only buyers in shared markets but also captive buyers.
While lowering the price can attract buyers from shared markets, doing so also reduces the revenue extracted from captive buyers who would have purchased at a higher price.
Thus, each seller should strategically set the price to balance the trade-off between these two opposite tensions.

\cite{BLN13} introduced the \emph{Bertrand networks} game to formalize such a strategic interaction. In their model, sellers correspond to vertices of an undirected graph, edges represent shared markets won by the cheaper endpoint, and vertex weights represent captive markets dedicated to a single seller. 
Every buyer has a maximum willingness to pay (valuation) of $1$, meaning that a buyer purchases the good as long as the lowest price offered to them is at most $1$, and sellers incur zero production cost. 
Although each seller simply chooses a scalar price in $[0,1]$, even a two-seller network generally lacks pure-strategy Nash equilibria and requires randomizing over continuous price intervals. Because price ties induce discontinuous payoffs, an equilibrium must be specified by a profile of probability distributions across the vertices.

While~\cite{BLN13} established the existence of equilibrium and designed a polynomial-time algorithm to compute an equilibrium for trees with a single captive market, the following fundamental questions on \emph{time and structural complexity} of equilibria have remained unresolved for the past decade:\footnote{Note that it has been known that mixed-strategy equilibria in Bertrand networks are continuous probability distributions over $[0,1]$~\citep{BLN13}. Accordingly,~\cite{BLN13} explicitly noted that whether one can write down the exact equilibrium with finite price intervals is a major open problem.}

\definecolor{mycolor}{rgb}{0.9, 0.90, 0.9}
\begin{tcolorbox}[colback=mycolor,colframe=gray!75!black,colframe=white]
\begin{center}
    \emph{
        Is there a polynomial-time algorithm to compute an equilibrium for any instance?
        \\
        Is the support of the equilibrium distributions necessarily finite for any instance?
    }
  \end{center}
\end{tcolorbox}

In this paper, we resolve these open problems, affirmatively and negatively. First, we settle the time complexity by proving that computing any approximate equilibrium within a small constant additive error is $\PPAD$-complete, even on bipartite graphs of maximum degree five with constant integer market weights. On the positive side, when both the maximum degree and the number of sellers with captive markets are bounded, we provide an efficient polynomial-time approximation scheme (EPTAS). 
Second, for exact equilibria, we show that any achievable profile of seller revenues can be realized by an equilibrium whose price distributions consist of at most $2^n$ intervals, on each of which every seller's cumulative distribution function (CDF) is linear in the reciprocal price $1/p$, thereby resolving the second open question affirmatively. We also prove a matching exponential lower bound, showing that there exist networks and achievable revenue profiles for which the unique equilibrium realizing those revenues requires exponentially many price intervals.

\subsection{Our Results}

To state our results formally, let $\alpha_i \geq 0$ denote the mass of captive buyers for seller $i$, and let $\beta_{ij} = \beta_{ji} > 0$ denote the mass of buyers in the shared market between adjacent sellers $i$ and $j$. Because prices lie in $[0,1]$ and seller $i$ can sell to at most $\alpha_i + \sum_j \beta_{ij}$ buyers, the total accessible buyer mass $M_i = \alpha_i + \sum_j \beta_{ij}$ is the maximum revenue seller $i$ can possibly earn. Rescaling seller $i$'s revenue by $1/M_i$ normalizes every seller's payoff to lie in $[0,1]$. Following the standard definition of additive approximate Nash equilibrium for games with normalized payoffs, we say that a profile of independent price distributions is a \emph{normalized $\eps$-equilibrium} if no seller $i$ can increase their expected revenue by more than $\eps M_i$ by unilaterally deviating to any price $p \in [0,1]$.\footnote{Measuring approximation error relative to $M_i$ is necessary for scale invariance: multiplying all market masses $\alpha_i$ and $\beta_{ij}$ by a positive constant rescales all revenues by the same factor without changing the strategic game. Moreover, unlike multiplicative approximation (which measures error relative to seller $i$'s equilibrium revenue $u_i$), normalizing by $M_i$ remains well-defined even when some sellers earn zero or arbitrarily small revenue in equilibrium.}
When we simply bound the right-hand side by $\eps$, we write \emph{unnormalized $\eps$-equilibrium} to distinguish it from the normalized version.
Throughout, we write $\eps$-equilibrium to denote normalized $\eps$-equilibrium for simplicity.

\paragraph{Complexity of approximate equilibrium}
Because mixed strategies in Bertrand networks are continuous distributions in general~\citep{BLN13}, formulating equilibrium computation as a finite search problem requires a finite representation of the output strategies. As we formalize in Section~\ref{sec:bh-membership}, a natural representation restricts the output distributions to a uniform price grid of step size $O(\eps)$, while still requiring the profile to be an $\eps$-equilibrium against \emph{all} continuous price deviations $p \in [0,1]$. Our first main result establishes that computing \emph{any} approximate equilibrium within a sufficiently small constant error is $\PPAD$-complete.\footnote{For computational problems, we assume the standard bit-complexity model in which all numerical inputs ($\alpha_i, \beta_{ij}$, and $\eps$) are rational numbers with total binary encoding length $\cL$.}

\begin{theorem}[Hardness of approximate equilibrium]
\label{thm:bh-hardness}
There exists a universal constant $\eps_0 > 0$ such that computing any $\eps_0$-equilibrium in a Bertrand network is $\PPAD$-complete. Moreover, $\PPAD$-hardness holds even on bipartite graphs of maximum degree five in which only degree-one vertices have captive buyers and all positive market masses belong to a fixed finite set of integers.
\end{theorem}

Note that this result rules out a polynomial-time approximation scheme (PTAS) on bounded-degree graphs unless $\PPAD = \mathsf{P}$. Notably, the hardness holds regardless of how the output distributions are represented: our reduction recovers an approximate Nash equilibrium of the underlying hard game from \emph{any} $\eps_0$-equilibrium of the Bertrand network, including arbitrary continuous price distributions, given that one can evaluate each seller's CDF at a single midpoint price to within a small constant accuracy.

In the hard instances of Theorem~\ref{thm:bh-hardness}, the maximum degree is five, but the number of sellers with captive buyers grows linearly with the size of the graph. As a complementary algorithmic result, we show that when both the maximum degree $\Delta$ and the number of sellers with captive buyers $k$ are bounded, the problem becomes tractable and admits an efficient polynomial-time approximation scheme (EPTAS).

\begin{theorem}[EPTAS for bounded degree and few captive sellers]
\label{thm:eptas-intro}
On any Bertrand network with maximum degree $\Delta$, at most $k$ sellers with positive captive mass, and input bit length $\cL$, for every $\eps \in (0,1]$ one can compute an $\eps$-equilibrium in time
\begin{align*}
 2^{O\rbr{k\eps^{-1}(\Delta+1)^{O(\eps^{-2})}}}\poly(\cL),
\end{align*}
where each seller's strategy is supported on at most $\Cel{8/\eps}$ prices.
\end{theorem}

For fixed $k$ and $\Delta$, the runtime of Theorem~\ref{thm:eptas-intro} depends only polynomially on the bit length $\cL$ of the market masses, independent of the ratio between the largest and smallest positive market masses.
We refer to Section~\ref{sec:eptas} (Algorithm~\ref{alg:eptas}) for more details.

\paragraph{Structural complexity of exact equilibria}
We next turn to the structural complexity of \emph{exact} equilibria ($\eps = 0$), \ie how many price intervals are needed to describe the sellers' continuous price distributions.
In an exact equilibrium, connected components of the graph without captive buyers yield zero revenue for all sellers and can be assigned price zero~\citep{BLN13}. On components with captive buyers, sellers randomize over continuous intervals in $(0,1)$, possibly with point masses (atoms) at the maximum price $p = 1$. \cite{BLN13} observed that if the active support intervals of an equilibrium are known, one can find an equilibrium in which, on each interval, every seller's CDF takes the form $F_i(p) = a_i + b_i/p$ for some constants $a_i$ and $b_i$. We call a strategy profile \emph{reciprocal-affine} if $(0,1)$ can be partitioned into finitely many intervals on each of which every seller's CDF is linear in $1/p$. 
On the other hand, it remains questionable whether every network admits a reciprocal-affine equilibrium with finitely many intervals, and how many intervals are needed, which we refer to as the \emph{structural complexity} of the equilibrium.

Our next main result resolves this question by showing that a common partition into at most $2^n$ intervals suffices to realize \emph{every} achievable equilibrium revenue vector $\bu = (u_1,\ldots,u_n)$.

\begin{theorem}[Finite exponential structural complexity]
\label{thm:finite}
For every Bertrand network on $n$ sellers and every achievable equilibrium revenue vector $\bu$, there exists an exact equilibrium with revenue vector $\bu$ whose marginal CDFs are reciprocal-affine on a common partition of $(0,1)$ into at most $2^n$ intervals (with possible atoms at price one, and price zero on connected components without captive buyers).
\end{theorem}

Notice that Theorem~\ref{thm:finite} holds for all graphs and market masses. Our proof also establishes two structural properties for any fixed positive revenue vector $\bu$: the set of equilibrium marginal CDF profiles realizing $\bu$ is convex (Corollary~\ref{cor:fiber-convex}), and the revenue-weighted aggregate CDF $\sum_{i=1}^n u_i F_i(p)$ is identical across all equilibria realizing $\bu$ (Corollary~\ref{cor:scalar-unique}). 
Furthermore, because Theorem~\ref{thm:finite} bounds the number of intervals by $2^n$, finding an exact equilibrium reduces to solving a finite system of polynomial equations and inequalities in $O(n2^n)$ variables (representing the interval breakpoints, CDF values, and revenues). Applying standard algorithms for solving polynomial systems~\citep{BPR96} yields an exact algorithm running in $2^{2^{O(n)}}\poly(\cL)$ time that computes an exact equilibrium (whose breakpoints are represented as roots of integer polynomials) and can also test whether a given revenue vector $\bu$ is achievable (Theorem~\ref{thm:exact-algorithm}).

Finally, we show that the exponential upper bound in Theorem~\ref{thm:finite} is tight when the target revenue vector $\bu$ is prescribed.

\begin{theorem}[Exponential lower bound on structural complexity]
\label{thm:exp-lower}
For every integer $k \geq 1$, one can construct in $\poly(k)$ time a Bertrand network on the complete graph of $N = 4k+4$ sellers with strictly positive captive and shared masses, together with an achievable revenue vector $\bu$, such that $\bu$ is realized by a unique equilibrium profile. Under this unique equilibrium, the union of the boundary points of the sellers' price supports contains at least $3\cdot 2^k - 3$ distinct prices in $(0,1)$, and therefore every common reciprocal-affine partition requires at least $3\cdot 2^k - 2$ intervals.
\end{theorem}

Together, Theorems~\ref{thm:finite} and~\ref{thm:exp-lower} establish that the structural complexity required to realize a prescribed equilibrium revenue vector is exponential in the number of sellers. We emphasize that Theorem~\ref{thm:exp-lower} fixes a target revenue vector $\bu$ and shows that the unique equilibrium achieving $\bu$ requires exponentially many intervals. Whether the same network admits a different equilibrium (with a different revenue vector) that has polynomial structural complexity remains an interesting open question.

\section{Overview of Techniques}
\label{sec:techniques}

\subsection{Binary Games via Continuous Prices}

Our hardness reduction starts from a two-action polymatrix game on a bipartite graph of maximum degree three, where computing a $1/20$-Nash equilibrium is proven to be $\PPAD$-hard~\citep{DFHM24}. In a two-action polymatrix game, each player $i$ chooses between two pure actions $s_i \in \cbr{0,1}$ and plays a separate $2 \times 2$ bimatrix game with each neighbor $j \in N(i)$, specified by a payoff matrix $A^{(i,j)} \in \R^{\cbr{0,1} \times \cbr{0,1}}$. A mixed strategy profile is given by a vector $\bx = (x_1, \ldots, x_n) \in [0,1]^n$, where $x_i$ denotes the probability that player $i$ plays action $1$ (and $1-x_i$ is the probability of playing action $0$). Given the neighbors' probabilities $\bx_{-i}$, player $i$'s expected payoff from playing pure action $s_i \in \cbr{0,1}$ is
\begin{align*}
 U_i(s_i, \bx_{-i}) = \sum_{j \in N(i)} \rbr{A^{(i,j)}_{s_i, 0}(1-x_j) + A^{(i,j)}_{s_i, 1} x_j},
\end{align*}
and player $i$'s expected payoff under their mixed strategy $x_i$ is $(1-x_i)U_i(0, \bx_{-i}) + x_i U_i(1, \bx_{-i})$. Define the expected \emph{payoff difference} between playing action $1$ and playing action $0$ by
\begin{align*}
 D_i(\bx) \defeq U_i(1, \bx_{-i}) - U_i(0, \bx_{-i}) = c_i + \sum_{j \in N(i)} d_{ij} x_j,
\end{align*}
where $c_i = \sum_{j \in N(i)} \rbr{A^{(i,j)}_{1,0} - A^{(i,j)}_{0,0}}$ and $d_{ij} = \rbr{A^{(i,j)}_{1,1} - A^{(i,j)}_{0,1}} - \rbr{A^{(i,j)}_{1,0} - A^{(i,j)}_{0,0}}$. Under the mixed strategy profile $\bx$, if player $i$ unilaterally deviates to pure action $a \in \cbr{0,1}$, their expected payoff increases by exactly $U_i(a, \bx_{-i}) - \sbr{(1-x_i)U_i(0, \bx_{-i}) + x_i U_i(1, \bx_{-i})} = (a - x_i) D_i(\bx)$. Thus, $\bx$ is a $\delta$-Nash equilibrium if and only if $(a - x_i) D_i(\bx) \leq \delta$ for every player $i$ and both pure actions $a \in \cbr{0,1}$. Note that because the two payoff matrices $A^{(i,j)}$ and $A^{(j,i)}$ on each edge $\cbr{i,j}$ are independent in a general polymatrix game, the two interaction coefficients $d_{ij}$ and $d_{ji}$ can be arbitrary real numbers with any signs and ratio.

To reduce this polymatrix game to a Bertrand network, a natural idea is to represent each player $i$ by a seller and encode their two actions $\cbr{0,1}$ as two target prices $a_i < b_i$ in $(0,1)$: posting the low price $a_i$ corresponds to playing action $0$, and posting the high price $b_i$ (with probability $x_i$) corresponds to playing action $1$. Turning this idea into a valid reduction requires overcoming three main obstacles:
\begin{enumerate}
\item \emph{A single shared market cannot realize arbitrary coefficient pairs $(d_{ij}, d_{ji})$:} In the polymatrix game, each edge $\cbr{i,j}$ has \emph{two} independent payoff coefficients $d_{ij}$ and $d_{ji}$, which can have any signs and any ratio. In a Bertrand network, however, a shared market on edge $\cbr{i,j}$ has only \emph{one} tunable parameter, its buyer mass $\beta_{ij} > 0$. As we show below, even when sellers only choose between two prices, a single shared market always makes one endpoint's payoff difference increase with the neighbor's probability while making the other's decrease in a fixed ratio. Thus, a single edge between two sellers cannot independently match both $d_{ij}$ and $d_{ji}$.
\item \emph{Splitting each player into multiple seller copies and keeping them synchronized:} Overcoming the limitation above requires combining multiple shared markets across different price menus and complementary probabilities ($x_i$ and $1-x_i$). Because a single seller can only use one price menu and one probability, we must split each player $i$ into multiple \emph{replica sellers} (which also allows us to bound every seller's degree by at most five, as claimed in Theorem~\ref{thm:bh-hardness}). However, all replicas of player $i$ must be forced to represent the same underlying probability $x_i$ in equilibrium. Enforcing this agreement requires connecting the replicas via internal markets of large mass $M_0$, which creates the risk that the $\delta$-equilibrium error gets multiplied by $M_0$ when we decode the players' payoffs.
\item \emph{Unrestricted continuous price deviations:} More importantly, in an actual Bertrand network, sellers are not restricted to a discrete two-price menu $\cbr{a_i, b_i}$, but can deviate to \emph{any} real price $p \in [0,1]$.
\end{enumerate}
We address the first two obstacles in an intermediate \emph{binary-menu game} where each seller is restricted to two prescribed prices $\cbr{a_i, b_i}$ and receives unary\footnote{A unary payoff term is only affected by the player's own action.} payoff offsets, and then resolve the third obstacle by attaching \emph{price-anchor} gadgets that enforce these two-price menus under unrestricted continuous pricing in $[0,1]$.

\paragraph{Spanning arbitrary payoff interactions via two edge orientations}
Consider first the intermediate binary-menu game where seller $i$ chooses between $\cbr{a_i, b_i}$ (playing $b_i$ with probability $x_i$) and seller $j$ chooses between $\cbr{a_j, b_j}$ (playing $b_j$ with probability $x_j$). For their price choices to interact strategically, their menus must be interleaved:
\begin{align*}
 a_i < a_j < b_i < b_j.
\end{align*}
Notice that this interleaving assigns asymmetric roles to the two endpoints: seller $i$ holds the lower price menu (which we call role $L$), and seller $j$ holds the higher price menu (which we call role $H$). On a shared market of mass $\beta > 0$ between role-$L$ seller $i$ and role-$H$ seller $j$, seller $i$ always wins the market when posting $a_i$ (earning $\beta a_i$) and wins when posting $b_i$ only if seller $j$ posts $b_j$ (earning $\beta b_i x_j$). Thus, this shared market contributes $\Delta_i = (\beta b_i) x_j - \beta a_i$ to seller $i$'s high-minus-low payoff difference. Conversely, seller $j$ never wins when posting $b_j$, and wins when posting $a_j$ only if seller $i$ posts $b_i$ (earning $\beta a_j x_i$), so the shared market contributes $\Delta_j = -(\beta a_j) x_i$ to seller $j$'s high-minus-low payoff difference.

Consequently, a single shared market with fixed roles ($i$ in role $L$ and $j$ in role $H$) always gives a positive linear coefficient $+\beta b_i$ to the role-$L$ seller and a negative linear coefficient $-\beta a_j$ to the role-$H$ seller, with a fixed ratio $-b_i/a_j$. A single shared market therefore cannot realize arbitrary target coefficients $(d_{ij}, d_{ji})$.

To realize any target pair $(d_{ij}, d_{ji}) \in \R^2$, we combine two ideas:
\begin{itemize}
\item \emph{Using both role orientations between $i$ and $j$:} For each source edge $\cbr{i,j}$, we give both players access to role-$L$ and role-$H$ copies. Connecting a role-$L$ copy of $i$ to a role-$H$ copy of $j$ (orientation $L_i$--$H_j$) produces a coefficient vector proportional to $(+b_L, -a_H)$, whereas connecting a role-$H$ copy of $i$ to a role-$L$ copy of $j$ (orientation $H_i$--$L_j$) produces the reversed coefficient vector $(-a_H, +b_L)$ after weighting the two roles appropriately. Because $(+b_L, -a_H)$ and $(-a_H, +b_L)$ are linearly independent, their linear combinations span all of $\R^2$.
\item \emph{Positive and negative replicas for nonnegative market masses:} Spanning $\R^2$ via linear combinations requires both positive and negative weights, whereas market masses $\beta$ must be nonnegative. To flip the sign of a coefficient while keeping $\beta \geq 0$, we introduce both \emph{positive replicas} of player $i$, whose high-price probability $x$ directly encodes player $i$'s action-$1$ probability $t_i$, and \emph{negative replicas} of player $i$, whose high-price probability $x$ encodes the complementary action-$0$ probability $1 - t_i$. Replacing $x_i = t_i$ by $x_i = 1 - t_i$ flips the sign of $t_i$'s coefficient from $-\beta a_H$ to $+\beta a_H$ while introducing a constant offset that we cancel using unary payoffs. Finally, rounding the source coefficients $c_i, d_{ij}$ to a fixed constant precision $\eta$ ensures that all resulting market masses come from a finite set.
\end{itemize}

\paragraph{Bounded-degree replication and weighted aggregation}
Because each player $i$ in the degree-three polymatrix game participates in at most three edges, and each edge can use both role orientations ($L$ and $H$) and both signs (positive and negative replicas), we create $12$ replicas per source player: three copies of each of the four role-sign pairs $L^+, H^+, L^-, H^-$. We assign at most one polymatrix edge to each replica. Together with at most two edges along the replica path and two price-anchor edges attached in the next step, this ensures that every replica has total degree at most $2 + 1 + 2 = 5$.

For the reduction to be valid, all $12$ replicas of player $i$ must be forced to represent the same underlying probability $t_i \in [0,1]$. Specifically, if we define the \emph{abstract probability} of replica $v$ as $t_v = x_v$ when $v$ is a positive replica and $t_v = 1 - x_v$ when $v$ is a negative replica, we need $t_v \approx t_w$ across all $12$ replicas. However, we cannot simply connect two replicas directly with an edge: as shown above, a single edge between two menus pushes the two endpoints' probabilities in the same direction rather than forcing them to be equal, and connecting replicas of the same player would also violate bipartiteness.

\begin{figure}[htbp]
\centering
\begin{tikzpicture}[x=1.03cm,y=0.92cm,
 replica/.style={circle,draw=blue!55!black,fill=blue!5,
  minimum size=7mm,inner sep=1pt,font=\small},
 aux/.style={rectangle,draw=black!65,fill=black!5,
  minimum size=4.7mm,inner sep=1pt,font=\scriptsize},
 market/.style={draw=black!65,line width=.65pt}]
\foreach \x/\lab/\name in {0/H^+_1/p1,2/L^+_1/p2,4/H^+_2/p3,
 6/L^+_2/p4,8/H^+_3/p5,10/L^+_3/p6}
 \node[replica] (\name) at (\x,0) {$\lab$};
\foreach \x/\lab/\name in {10/L^-_1/n1,8/H^-_1/n2,6/L^-_2/n3,
 4/H^-_2/n4,2/L^-_3/n5,0/H^-_3/n6}
 \node[replica,draw=violet!65!black,fill=violet!5] (\name) at (\x,-2) {$\lab$};
\foreach \x/\j in {1/1,3/2,5/3,7/4,9/5}{
 \node[aux] (m\j) at (\x,0) {$M$};
 \node[aux] (v\j) at (\x,-2) {$M$};}
\foreach \left/\mid/\right in {p1/m1/p2,p2/m2/p3,p3/m3/p4,p4/m4/p5,p5/m5/p6,
 n6/v1/n5,n5/v2/n4,n4/v3/n3,n3/v4/n2,n2/v5/n1}{
 \draw[market] (\left)--(\mid)--(\right);}
\node[aux] (z) at (10,-1) {$Z$};
\draw[market] (p6)--(z)--(n1);
\node[font=\small,align=center] at (4.5,-1)
 {All $12$ replicas represent the same abstract probability.};
\draw[market,dashed] (p3.north)--(4,.8);
\node[font=\small,anchor=south] at (4,.86) {At most one source edge per replica};
\node[font=\small,anchor=north] at (4.5,-2.55)
 {Two price anchors are later attached to every replica and auxiliary.};
\end{tikzpicture}
\caption{Bounded-degree replica path for a single source player. Circular
nodes represent replicas and square nodes represent auxiliary sellers.
Each equality link passes through an $M$ auxiliary, while the $Z$ link
couples positive and negative replicas. Every replica has degree at most
five in the final graph, consisting of at most two edges along the
replica path, at most one polymatrix source edge, and two price-anchor
edges.}
\label{fig:replicas}
\end{figure}
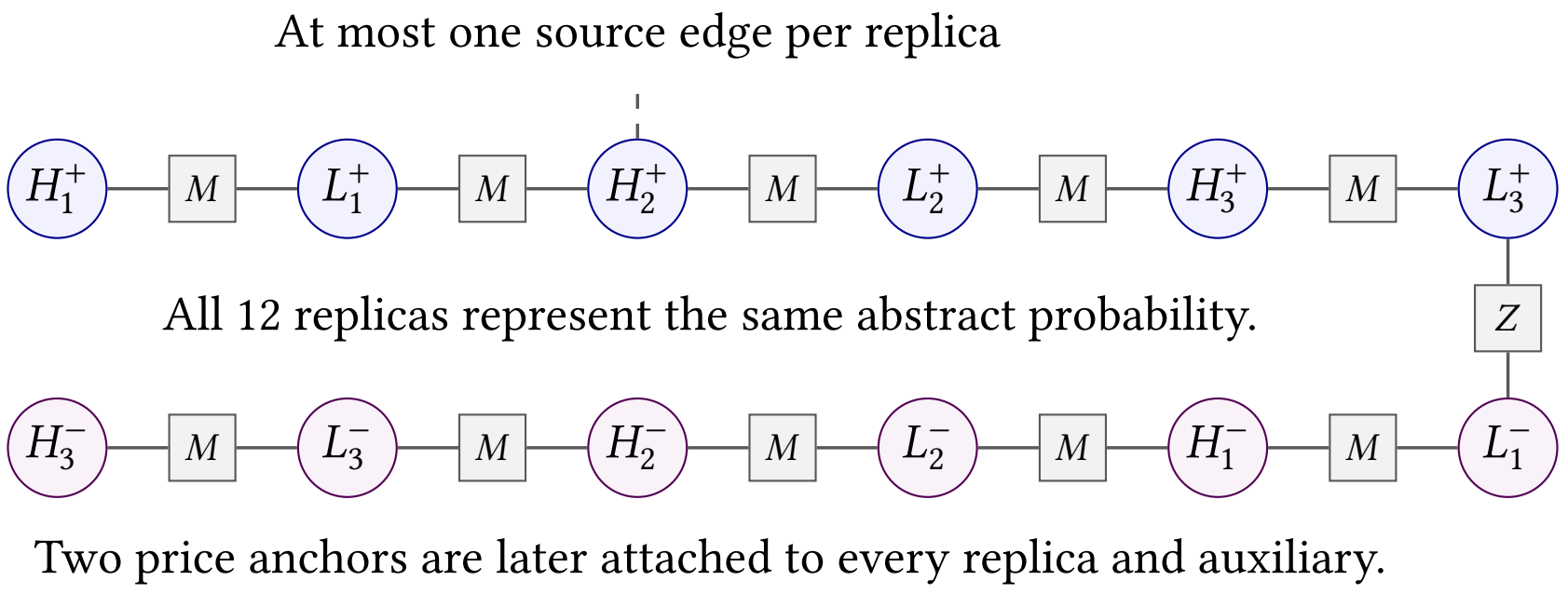

Instead, we connect the $12$ replicas into a path by placing an \emph{auxiliary seller} between each consecutive pair of replicas, as shown in Figure~\ref{fig:replicas}:
\begin{itemize}
\item \emph{Equality links via middle-menu ($M$) auxiliaries:} Within the six positive replicas (and similarly within the six negative replicas), replicas alternate between role $L$ and role $H$. To synchronize an adjacent pair of role-$L$ and role-$H$ replicas, we place an auxiliary seller between them whose two-price menu $M = \cbr{a_M, b_M}$ lies strictly \emph{between} menu $L$ and menu $H$, \ie $a_L < a_M < a_H < b_L < b_M < b_H$. Because menu $M$ is higher than $L$ but lower than $H$, the $L$-replica's probability $x_L$ decreases $M$'s payoff difference while the $H$-replica's probability $x_H$ increases it. Scaling both edges by a large mass $M_0$ makes the $M$-auxiliary's payoff difference proportional to $M_0(x_H - x_L)$, while the $M$-auxiliary's probability $y_M$ pushes $x_L$ upward and $x_H$ downward. This creates a restoring feedback loop: whenever $x_H > x_L$, the $M$-auxiliary plays $y_M \approx 1$, pushing $x_L$ up and $x_H$ down until $x_L \approx x_H$.
\item \emph{Complementation link via a high-menu ($Z$) auxiliary:} To couple the positive replica path to the negative replica path, we connect the last positive replica $L^+_3$ and the first negative replica $L^-_1$ through an auxiliary seller whose two-price menu $Z = \cbr{a_Z, b_Z}$ lies strictly above menu $L$, \ie $a_L < a_Z < b_L < b_Z$. Because both $L^+_3$ and $L^-_1$ have lower menus than $Z$, both $x_{L^+_3}$ and $x_{L^-_1}$ decrease $Z$'s payoff difference. Adding an offset of $M_0$ makes $Z$'s payoff difference proportional to $M_0(1 - x_{L^+_3} - x_{L^-_1}) = M_0(t_{L^-_1} - t_{L^+_3})$, which analogously forces $x_{L^-_1} \approx 1 - x_{L^+_3}$, or equivalently $t_{L^+_3} \approx t_{L^-_1}$.
\end{itemize}

For each internal link $e=(v,w)$ along the replica path (oriented from its $L$-replica $v$ to its $H$-replica $w$, or from $v = L^+_3$ to $w = L^-_1$ on the $Z$-link), let $y_e \in [0,1]$ denote the auxiliary seller's high-price probability (or one minus that probability on negative equality links). Under this orientation, the auxiliary seller on link $e=(v,w)$ faces an abstract payoff difference of $M_0(t_w - t_v)$. Furthermore, by weighting the role-$L$ and role-$H$ payoffs appropriately ($w_L = 1$ and $w_H = 24/5$), link $e=(v,w)$ contributes equal and opposite weighted payoff signals $+k_e(y_e - 1/2)$ to replica $v$ and $-k_e(y_e - 1/2)$ to replica $w$, where $k_e = \Theta(M_0)$. This cancellation property yields two key guarantees in any unnormalized $\delta$-equilibrium:
\begin{itemize}
\item \emph{Approximate consensus across all $12$ replicas:} If the $12$ abstract probabilities $t_v$ were not close to one another, we could partition the replicas into a group $S$ with higher probabilities and its complement with lower probabilities across a gap of size $\tau$. Summing the weighted payoffs over all replicas in $S$, the signals on links internal to $S$ cancel in pairs, whereas any link crossing out of $S$ has $\abs{t_w - t_v} > \tau$ and exerts a strong downward payoff pressure of $-\Omega(M_0)$ on $S$. For sufficiently large constant $M_0$, this downward pressure overwhelms the bounded external source payoffs and contradicts approximate best response for replicas in $S$, proving that all $12$ probabilities $t_v$ lie within $O(\delta/M_0)$ of one another.
\item \emph{Telescoping internal signals without error amplification:} Even though all $t_v$'s are within $O(\delta/M_0)$ of a common value $t_i$, an individual replica's payoff is dominated by the large consensus mass $M_0$, so its individual best-response inequality has an error of order $M_0 \cdot O(\delta/M_0)$ if we naively substitute $t_i$. Instead, we sum the weighted best-response inequalities $(a - t_v) w_v g_v \leq w_v \delta$ across all $12$ replicas of player $i$. Because link $e=(v,w)$ contributes $+k_e(y_e - 1/2)$ to $w_v g_v$ and $-k_e(y_e - 1/2)$ to $w_w g_w$, multiplying by $(a - t_v)$ and $(a - t_w)$ respectively and summing gives
\begin{align*}
 (a - t_v)k_e(y_e - 1/2) - (a - t_w)k_e(y_e - 1/2) = k_e(t_w - t_v)(y_e - 1/2).
\end{align*}
At the same time, the auxiliary seller on link $e$ has payoff difference $M_0(t_w - t_v)$ and is itself in an unnormalized $\delta$-equilibrium. Therefore, if the auxiliary seller deviates from its probability $y_e$ to the uniform $(1/2, 1/2)$ mixture, its expected payoff cannot increase by more than $\delta$, which means $(1/2 - y_e) M_0(t_w - t_v) \leq \delta$, or equivalently $M_0(t_w - t_v)(y_e - 1/2) \geq -\delta$. Since $k_e = \Theta(M_0)$ is a positive constant multiple of $M_0$, each internal term $k_e(t_w - t_v)(y_e - 1/2)$ is bounded from below by $-O(\delta)$. Moving these internal terms to the right-hand side therefore increases the error by at most $O(\delta)$ rather than $O(M_0)$, leaving only the aggregated external polymatrix payoff difference $(a - t_i) D_i(\bt) \leq O(\delta)$.
\end{itemize}

\paragraph{Enforcing two-price menus via price anchors}
The construction so far assumes that each seller $i$ in the binary-menu game (namely, each of the $12$ replica sellers and $11$ auxiliary sellers per player, which we call \emph{computational sellers} to distinguish them from the anchor sellers introduced below) only chooses between two menu prices $\cbr{a_i, b_i}$ and receives unary payoff offsets $(d_{i,0}, d_{i,1})$. In an actual Bertrand network, however, every seller can post any real price $p \in [0,1]$ and there are no exogenous unary rewards.

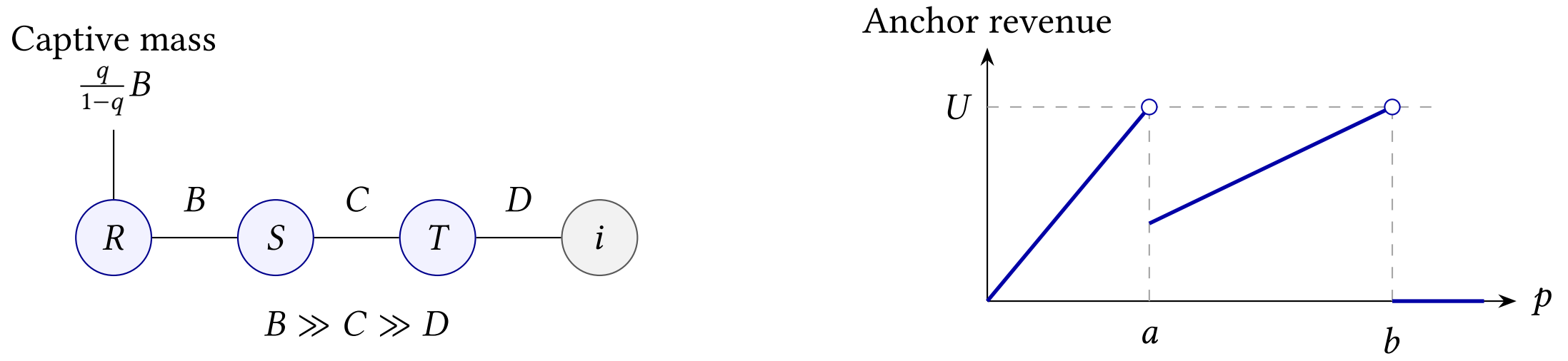
\begin{figure}[htbp]
\centering
\begin{minipage}[b]{.49\linewidth}
\centering
\begin{tikzpicture}[x=1cm,y=1cm,
 seller/.style={circle,draw=blue!55!black,fill=blue!5,
 minimum size=7mm,inner sep=1pt,font=\small}]
\node[seller] (r) at (0,0) {$R$};
\node[seller] (s) at (1.5,0) {$S$};
\node[seller] (t) at (3,0) {$T$};
\node[seller,fill=black!5,draw=black!65] (i) at (4.5,0) {$i$};
\draw (r)--node[above=3pt,font=\small] {$B$}(s);
\draw (s)--node[above=3pt,font=\small] {$C$}(t);
\draw (t)--node[above=3pt,font=\small] {$D$}(i);
\node[align=center,font=\small,anchor=south] (cap) at (0,1)
 {Captive mass\\$\frac{q}{1-q}B$};
\draw (cap.south)--(r.north);
\node[font=\small] at (2.25,-.8) {$B\gg C\gg D$};
\node[font=\small,align=center] at (2.25,-1.35)
 {Terminal $T$ concentrates near $q$.};
\end{tikzpicture}
\end{minipage}\hfill
\begin{minipage}[b]{.49\linewidth}
\centering
\begin{tikzpicture}[x=1cm,y=1cm,>=Stealth]
\draw[->] (0,0)--(4.9,0) node[right,font=\small] {$p$};
\draw[->] (0,0)--(0,2.35) node[above,font=\small] {Anchor revenue};
\draw[dashed,black!35] (0,1.8)--(4.15,1.8);
\node[left,font=\small] at (0,1.8) {$U$};
\draw[dashed,black!35] (1.5,0)--(1.5,1.8);
\draw[dashed,black!35] (3.75,0)--(3.75,1.8);
\draw[blue!65!black,line width=1pt] (0,0)--(1.5,1.8);
\draw[blue!65!black,line width=1pt] (1.5,.72)--(3.75,1.8);
\draw[blue!65!black,line width=1pt] (3.75,0)--(4.6,0);
\draw[blue!65!black,fill=white] (1.5,1.8) circle (2pt);
\draw[blue!65!black,fill=white] (3.75,1.8) circle (2pt);
\node[below=3pt,font=\small] at (1.5,0) {$a$};
\node[below=3pt,font=\small] at (3.75,0) {$b$};
\node[font=\small,align=center] at (2.5,-.8)
 {Two menu actions share the same baseline.};
\end{tikzpicture}
\end{minipage}
\caption{Left: a three-seller price anchor $R$--$S$--$T$ attached to computational
seller $i$. Right: seller $i$'s combined expected revenue from its two attached
price anchors (one tuned to $q = a$ and one tuned to $q = b$) as a function of
seller $i$'s price $p$, exhibiting two peaks at $a$ and $b$.}
\label{fig:anchor-menu}
\end{figure}

To enforce the two-price menus and unary payoffs through pure price competition, we first construct a three-seller \emph{price anchor} path $R$--$S$--$T$ that forces its terminal seller $T$ to concentrate its price distribution near a prescribed target price $q \in (0,1)$ (Figure~\ref{fig:anchor-menu}, left). In this gadget, the leaf seller $R$ has a captive market of mass $\frac{q}{1-q}B$, the shared market $(R,S)$ has a large mass $B$, and the shared market $(S,T)$ has a much smaller mass $C \ll B$. To see why $T$ concentrates near $q$, consider the exact duopoly between $R$ and $S$ (ignoring the tiny edge $C \ll B$). By posting price $1$, seller $R$ can always guarantee revenue $\frac{q}{1-q}B$ from its captive buyers, which equals the maximum revenue $R$ could get at price $q$ even if $R$ won the entire shared market $B$ (since $q\rbr{\frac{q}{1-q}B + B} = \frac{q}{1-q}B$). Thus, in the exact $R$--$S$ duopoly~\citep{BLN13}, neither $R$ nor $S$ prices below $q$, and for seller $R$ to be indifferent across all prices $p \in [q,1]$, seller $S$'s survival probability $Z(p) = \Pr{P_S \geq p}$ must satisfy $p\rbr{\frac{q}{1-q}B + B Z(p)} = \frac{q}{1-q}B$, which yields
\begin{align*}
 Z(p) = \Pr{P_S \geq p} = \min\rbr{1, \frac{q}{1-q}\rbr{\frac{1}{p} - 1}}.
\end{align*}
In particular, $Z(p) = 1$ for all $p \in [0,q]$ and $Z(p) = \frac{q}{1-q}(1/p - 1)$ for $p \in [q,1]$. Consequently, when terminal seller $T$ competes against $S$ on the small edge $(S,T)$ of mass $C$, $T$'s expected revenue at price $p$ is $C p Z(p)$, which increases linearly as $C p$ for $p \in [0,q]$, reaches a sharp unique maximum of $C q$ at $p = q$, and decreases linearly as $C\frac{q}{1-q}(1-p)$ for $p \in [q,1]$. Using integral envelopes of the survival functions, we prove that this sharp revenue peak at $p = q$ is robust to $\eps$-equilibrium errors and weak external edges of mass $D \ll C$ incident to $T$, forcing $\Ex{\abs{P_T - q}}$ to be arbitrarily small.

To enforce a two-price menu $\cbr{a_i, b_i}$ at computational seller $i$, we take two separate copies of this three-seller price-anchor gadget: one anchor $R_{a_i}$--$S_{a_i}$--$T_{a_i}$ tuned to target price $q = a_i$ (connecting $T_{a_i}$ to seller $i$ with edge mass $W_0$), and a second anchor $R_{b_i}$--$S_{b_i}$--$T_{b_i}$ tuned to target price $q = b_i$ (connecting $T_{b_i}$ to seller $i$ with edge mass $W_1$). Figure~\ref{fig:anchor-menu} (right) plots seller $i$'s combined expected revenue from these two anchor edges as a function of seller $i$'s own price $p_i \in [0,1]$ in the ideal limit where $T_{a_i}$ prices at $a_i$ and $T_{b_i}$ prices at $b_i$:
\begin{itemize}
\item For $p_i < a_i$, seller $i$ undercuts both terminals $T_{a_i}$ and $T_{b_i}$ and wins both anchor markets, so its anchor revenue increases linearly as $(W_0 + W_1)p_i$, approaching $(W_0 + W_1)a_i$ as $p_i \uparrow a_i$.
\item Once $p_i > a_i$, seller $i$ loses the market with $T_{a_i}$ (causing the downward jump at $p = a_i$ in Figure~\ref{fig:anchor-menu}, right). For $p_i \in (a_i, b_i)$, seller $i$ wins only the market with $T_{b_i}$, so its anchor revenue increases linearly again as $W_1 p_i$, approaching a second peak of $W_1 b_i$ as $p_i \uparrow b_i$.
\item For $p_i > b_i$, seller $i$ loses both anchor markets and earns zero anchor revenue.
\end{itemize}
By setting the attachment masses $W_0$ and $W_1$ so that $(W_0 + W_1)a_i = U + d_{i,0}$ and $W_1 b_i = U + d_{i,1}$ for a large common baseline $U$, seller $i$'s combined anchor revenue has two sharp local maxima at $p_i \approx a_i$ (of height $U + d_{i,0}$) and $p_i \approx b_i$ (of height $U + d_{i,1}$). Any price outside small $\sigma$-neighborhoods of $a_i$ and $b_i$ loses at least $\Omega(U\sigma)$ in anchor revenue, which forces seller $i$ to place nearly all probability near $a_i$ and $b_i$ while simultaneously realizing the desired unary payoffs $(d_{i,0}, d_{i,1})$. Moreover, because both peaks share the same baseline $U$, comparing seller $i$'s revenue pointwise against $U + d_{i,A_i(p)}$ cancels $U$ out of the best-response inequality so that off-menu probability is penalized only by the bounded strategic scale $E$ rather than $U$.

Combining these ingredients completes the hardness reduction of Theorem~\ref{thm:bh-hardness}. Suppose we are given any normalized $\eps_0$-equilibrium of the constructed Bertrand network. Because every computational seller $i$ has total accessible market mass $M_i$ bounded by a fixed constant $E + 10U \leq 11U = O(1)$ independent of the network size, its normalized deviation bound $\eps_0 M_i$ immediately implies an unnormalized deviation bound of $O(\eps_0)$ in the continuous game. The price anchors then force each computational seller $i$ to concentrate its price distribution near its two menu prices $\cbr{a_i, b_i}$, so thresholding each seller's CDF at the midpoint $(a_i + b_i)/2$ yields an unnormalized $\delta$-equilibrium of the intermediate binary-menu game with $\delta = O(\eps_0)$. Finally, aggregating the deviation inequalities over the $12$ replicas of each source player cancels the internal consensus signals and bounds each player's maximum deviation gain in the polymatrix game by $O(\delta) = O(\eps_0) < 1/20$. Since the source polymatrix game already has total payoffs normalized to $[0,1]$, this directly recovers a $1/20$-Nash equilibrium.


\paragraph{$\PPAD$ membership and bounded-captive EPTAS}
For $\PPAD$ membership, restricting all sellers to a uniform price grid of step size $h = O(\eps)$ yields a finite polymatrix game whose exact Nash equilibrium lies in $\PPAD$. 
Any continuous price deviation $p \in [0,1]$ can be rounded down to the nearest grid point $p^-$ without losing any shared market already won at $p$, while reducing the price by at most $h$. Thus, an exact equilibrium of the finite grid game is automatically an $\eps$-equilibrium of the continuous Bertrand network against all real-price deviations.

When both the maximum degree $\Delta$ and the number of captive sellers $k$ are bounded, we obtain an EPTAS by exploiting scale invariance away from captive markets: for any seller without captive buyers, multiplying both its own price and all its neighbors' prices by a common factor $\lambda \in [0,1]$ scales both its actual and best-response revenues by $\lambda$. We solve a finite grid game inside a radius-$R$ neighborhood around the $k$ captive sellers, smooth each positive grid atom into a uniform distribution over the adjacent grid cell of width $h$, and linearly scale prices down to zero via $\lambda_i = 1 - d_i/R$ as a function of the graph distance $d_i$ from the captives. Because uniform smoothing bounds the total variation distance between adjacent scaling factors $\lambda_i$ and $\lambda_j$ by $O(1/(Rh))$, sellers inside the neighborhood experience negligible error from the gradual scale attenuation, and sellers at distance $d_i \geq R$ can simply post price zero.
\subsection{Duality and Polyhedral Replay}

To prove our $2^n$ upper bound on the structural complexity of exact equilibria (Theorem~\ref{thm:finite}), we fix an achievable positive revenue vector $\bu$ after assigning price zero to connected components without captive buyers. Changing variables to reciprocal price $t = 1/p - 1$ and writing $U_i(t) = \Pr{P_i \geq (1+t)^{-1}}$, the condition that seller $i$ cannot earn more than $u_i$ by undercutting price $p = (1+t)^{-1}$ becomes the linear inequality
\begin{align*}
 B\bU(t) \leq \bu(1+t) - \balpha,
\end{align*}
where $B$ is the symmetric weighted adjacency matrix of shared markets. While this inequality ensures that no seller can earn \emph{more} than $u_i$, we also need each seller's expected revenue to actually \emph{equal} $u_i$. By counting the total buyer mass $\cA$ purchased across the entire network, we show that for any profile $\bU$ satisfying the linear inequalities above,
\begin{align*}
 \sum_i u_i \Ex{P_i^{-1}} - \cA
 = \sum_i \Ex{\frac{u_i - P_i d_i(P_i)}{P_i}} \geq 0,
\end{align*}
where $d_i(p)$ is seller $i$'s expected demand at price $p$. Equality holds if and only if $P_i d_i(P_i) = u_i$ almost surely for every seller $i$. Since integration by parts rewrites $\sum_i u_i \Ex{P_i^{-1}}$ as a constant minus $\int_0^T \bu^\top \bU(t)\,dt$, the exact equilibria realizing revenue vector $\bu$ are precisely the profiles $\bU$ that maximize the linear integral objective $\int_0^T \bu^\top \bU(t)\,dt$ subject to $B\bU(t) \leq \bu(1+t) - \balpha$ and monotonicity. In effect, this deficit identity acts as a continuous linear programming \emph{duality} relation: $\sum_i u_i \Ex{P_i^{-1}}$ is bounded below by the total buyer mass $\cA$, and zero duality gap holds if and only if $\bU$ satisfies complementary slackness against the linear undercut constraints.

This continuous linear programming formulation allows us to adapt the duality and complementary slackness techniques of~\cite{SW14,SW15}. Because linearly interpolating $\bU(t)$ between its endpoint values $\bU(a)$ and $\bU(b)$ on any subinterval $[a,b]$ preserves all linear constraints, maximizing $\int_0^T \bu^\top \bU(t)\,dt$ forces the scalar function $V(t) = \bu^\top \bU(t)$ to be concave. At almost every $t$, the subset of active sellers $S = \cbr{i : U_i'(t) > 0}$ must have zero slack in their revenue constraints by complementary slackness. Letting $B_{SS}$ denote the $|S| \times |S|$ principal submatrix of $B$ restricted to sellers in $S$ (and $\bU_S, \bu_S, \balpha_S$ the corresponding subvectors), differentiating the active constraints $B_{SS}\bU_S(t) = \bu_S(1+t) - \balpha_S$ yields $B_{SS}\bU_S'(t) = \bu_S$. Even if $B_{SS}$ is singular, symmetry of $B_{SS}$ guarantees that the scalar slope $V'(t) = \bu_S^\top \bU_S'(t)$ takes a unique value for each subset $S \subseteq [n]$. Since $V$ is concave, its nonincreasing derivative $V'(t)$ can take each of the at most $2^n$ possible slope values on at most one interval. Interpolating each seller's marginal $U_i$ linearly across each of these at most $2^n$ intervals preserves both the constraints and the scalar objective $V(t)$, yielding a reciprocal-affine equilibrium with structural complexity at most $2^n$.

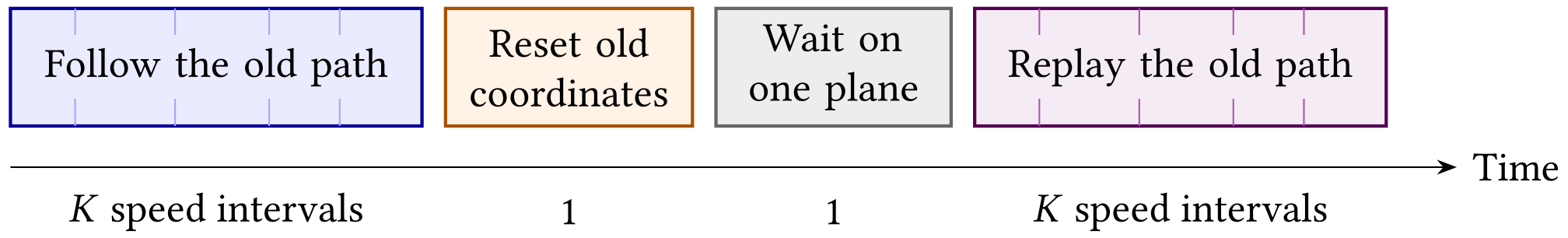
\begin{figure}[htbp]
\centering
\begin{tikzpicture}[x=1cm,y=1cm,>=Stealth,font=\small]
\draw[->] (0,0)--(12.3,0) node[right] {Time};
\fill[blue!8] (0,.35) rectangle (3.5,1.35);
\fill[orange!10] (3.7,.35) rectangle (5.8,1.35);
\fill[black!7] (6,.35) rectangle (8,1.35);
\fill[violet!8] (8.2,.35) rectangle (11.7,1.35);
\draw[blue!60!black,line width=.8pt] (0,.35) rectangle (3.5,1.35);
\draw[orange!65!black,line width=.8pt] (3.7,.35) rectangle (5.8,1.35);
\draw[black!60,line width=.8pt] (6,.35) rectangle (8,1.35);
\draw[violet!65!black,line width=.8pt] (8.2,.35) rectangle (11.7,1.35);
\foreach \x in {.55,1.4,2.2,2.8} \draw[blue!35] (\x,.35)--(\x,1.35);
\foreach \x in {8.75,9.6,10.4,11} \draw[violet!65] (\x,.35)--(\x,1.35);
\node[fill=blue!8,inner sep=3pt,align=center,text width=3.1cm]
 at (1.75,.85) {Follow the old path};
\node[align=center,text width=1.85cm] at (4.75,.85) {Reset old\\coordinates};
\node[align=center,text width=1.75cm] at (7,.85) {Wait on\\one plane};
\node[fill=violet!8,inner sep=3pt,align=center,text width=3.1cm]
 at (9.95,.85) {Replay the old path};
\node at (1.75,-.38) {$K$ speed intervals};
\node at (4.75,-.38) {$1$};
\node at (7,-.38) {$1$};
\node at (9.95,-.38) {$K$ speed intervals};
\end{tikzpicture}
\caption{One inductive step of the polyhedral ascent construction,
replacing $K$ speed intervals by $2K+2$. Two auxiliary coordinates allow
the trajectory to reset the previous coordinates to zero and replay the
entire path at a slower speed.}
\label{fig:replay}
\end{figure}

To prove the matching exponential lower bound on structural complexity (Theorem~\ref{thm:exp-lower}), we reverse this connection between active sets and concave slopes via a recursive \emph{polyhedral replay} construction. We inductively construct a concave polyhedral function $f_k(\by) = \min_i (\bp_i^\top \by + b_i)$ on $2k+1$ affine planes whose steepest-ascent trajectory $\by(t)$, governed by the time derivative $\frac{d\by(t)}{dt} = \bg(\by(t))$, passes through $K_k = 3\cdot 2^k - 2$ distinct speed intervals. To generate exponentially many speed intervals using only $O(k)$ planes, each inductive step adds two new coordinates and two new planes that cause the trajectory to first follow the $(k-1)$-st polyhedral path of $K_{k-1}$ pieces, then move along the first new plane to reset all previous coordinates back to zero, wait on the second new plane, and finally \emph{replay} (traverse a second time) the entire $(k-1)$-st polyhedral path at a slower speed, producing $2K_{k-1} + 2$ distinct speed intervals in total (Figure~\ref{fig:replay}). Explicit lower bounds on the length and speed separation of every interval ensure that rounding the plane coefficients to polynomial bit length preserves all $3\cdot 2^k - 2$ pieces, and closed-form recurrences for the trajectory endpoints allow us to compute these coefficients in $\poly(k)$ time without simulating the exponentially many steps. Finally, integrating the active convex combination weights along the trajectory and subtracting their positive-definite Gram matrix $D$ from a large rank-one matrix $\rho \bone \bone^\top$ embeds this trajectory into a twin Bertrand network on $N = 4k+4$ sellers whose market matrix $B$ is strictly negative definite on $\bu^\perp$. Negative definiteness on $\bu^\perp$ forces the equilibrium realizing $\bu$ to be unique, so its CDFs must undergo all $3\cdot 2^k - 2$ slope changes.

\section{Further Related Works}

\paragraph{Bertrand networks}
\cite{BLN13} introduced the Bertrand network model, established equilibrium existence via limit arguments, and characterized structural properties of mixed-strategy supports. They designed a polynomial-time algorithm for trees with a single captive market and introduced the notion of a finite \emph{sketch} consisting of support intervals, boundary points, and atoms at price one. Given a valid sketch, their Theorem~5.8 reconstructs a reciprocal-affine equilibrium via linear programming. In Section~9 of their work, they asked whether equilibria can be computed efficiently on general networks and whether finite sketches always suffice. Theorem~\ref{thm:bh-hardness} resolves their first question negatively even for approximate equilibria with a fixed constant error, while Theorem~\ref{thm:finite} resolves their second question by establishing a tight $2^n$ upper bound on the number of intervals needed to realize any achievable revenue vector.

\paragraph{Computational complexity of equilibrium}
Our $\PPAD$-hardness reduction builds on the hardness of finding constant-error equilibria in sparse polymatrix games. \cite{Rubinstein18} first established constant-error $\PPAD$-hardness for two-action, degree-three polymatrix games, and~\cite{DFHM24} sharpened the quantitative inapproximability threshold via the \emph{Pure-Circuit} problem. For $\PPAD$ membership, discretizing the price space to a uniform grid yields a finite polymatrix game whose exact equilibrium problem lies in $\PPAD$ by~\cite[Corollary~25(3)]{EY10}, and a downward price-rounding argument controls off-grid deviations despite tie-breaking discontinuities. In a related vendor pricing model,~\cite{CCKKPV17} studied pure equilibria and subsidies while leaving the complexity of mixed equilibria open.

\paragraph{Structural complexity and continuous linear programming}
Our proof of Theorem~\ref{thm:finite} draws on structural techniques from continuous linear programming with constant coefficients. \cite{SW14,SW15} demonstrated that optimal cumulative controls admit piecewise-linear representations by analyzing the concavity of a scalar projection, bounding the set of admissible scalar slopes, and linearly interpolating between interval endpoints. We adapt this approach to Bertrand networks by deriving a buyer-counting identity that equates equilibrium conditions at fixed revenues to maximizing a linear integral objective, and by using the symmetry of the market matrix to show that each active seller subset induces at most one scalar slope. Combining our $2^n$ interval bound with standard algorithms for solving polynomial systems~\citep{BPR96,BPR06} yields our doubly exponential exact algorithm.

\paragraph{Price competition with consideration sets and networks}
The economic trade-off between captive and contested buyers also appears in oligopoly models with consumer consideration sets. \cite{AV22} characterized equilibrium pricing under structured consideration patterns, \cite{JR21} analyzed equilibrium uniqueness under symmetric consideration, and \cite{MR26} studied asymmetric captive-and-shopper markets with heterogeneous costs. In the algorithmic game theory literature, alternative network pricing models place sellers on capacitated edges with path buyers~\citep{CR08,AS15}, model edges as complementary goods requiring purchase from both endpoints~\citep{BBN16}, consider combinatorial buyers or inventory dynamics~\citep{BNP14,BPS15}, or study smooth network externalities~\citep{Hinnosaar25}. Unlike these models, each shared market in a Bertrand network is won by the cheaper of two substitute sellers.

\section{Model and Preliminaries}
\label{sec:model}

A Bertrand network is defined by a finite simple undirected graph $G=(V,E)$ on $n=\abs{V}$ vertices, nonnegative captive market masses $\alpha_i \geq 0$ for each vertex $i \in V$, and positive shared market masses $\beta_{ij}=\beta_{ji}>0$ for each edge $\cbr{i,j} \in E$ (with $\beta_{ij}=0$ for $\cbr{i,j}\notin E$, including $i=j$).
Each seller $i \in V$ chooses a price $p_i \in [0,1]$. Captive buyers at vertex $i$ purchase from seller $i$ at any price $p_i \in [0,1]$. Buyers on an edge $\cbr{i,j} \in E$ purchase from the cheaper endpoint and split equally if $p_i=p_j$. Given a pure price profile $\bp = (p_1,\ldots,p_n) \in [0,1]^V$, seller $i$'s revenue is
\begin{align}
 \pi_i(\bp)=p_i\rbr{\alpha_i+
 \sum_{j:\cbr{i,j}\in E}\beta_{ij}
 \rbr{\Ind{p_i<p_j}+\tfrac12\Ind{p_i=p_j}}}.
 \label{eq:model-payoff}
\end{align}
Isolated vertices with $M_i \defeq \alpha_i + \sum_{j:\cbr{i,j}\in E}\beta_{ij} = 0$ earn zero revenue under all price profiles and are assigned price zero.

\paragraph{Mixed strategies and equilibrium}
A mixed strategy for seller $i$ is an independent probability measure on $[0,1]$, represented by a random price $P_i$ with right-continuous CDF $F_i(p)=\Pr{P_i\leq p}$. Given a mixed strategy profile $\bF = (F_1,\ldots,F_n)$, we write $u_i(p,\bF_{-i})=\Ex{\pi_i(p,\bP_{-i})}$ for seller $i$'s expected revenue when posting price $p \in [0,1]$ and $u_i(\bF)=\Ex{\pi_i(\bP)}$ for seller $i$'s expected revenue under $\bF$.

A profile $\bF$ is a \emph{normalized $\eps$-equilibrium} if no seller $i \in V$ can increase their expected revenue by more than $\eps M_i$ via a unilateral deviation:
\begin{align}
 \sup_{p\in[0,1]}u_i(p,\bF_{-i})-u_i(\bF)\leq\eps M_i,
 \label{eq:model-equilibrium}
\end{align}
where $M_i=\alpha_i+\sum_j\beta_{ij}$ is seller $i$'s total accessible market mass. We refer to an unnormalized bound $\sup_{p\in[0,1]}u_i(p,\bF_{-i})-u_i(\bF)\leq\delta$ as an \emph{unnormalized $\delta$-equilibrium}, and an \emph{exact equilibrium} corresponds to $\eps=0$ for all $i \in V$. Taking the supremum in~\eqref{eq:model-equilibrium} accounts for the fact that undercutting an opponent's point mass (atom) yields a limiting payoff as the price approaches the atom from below that need not be attained at any single price. For ease of exposition, we write $\eps$-equilibrium to denote a normalized $\eps$-equilibrium unless specified otherwise.

\paragraph{Reciprocal-affine representations and structural complexity}
A revenue vector $\bu = (u_1,\ldots,u_n)$ is \emph{achievable} if it is induced by some exact equilibrium $\bF$, which always exists by~\cite[Theorem~4.3]{BLN13}. Further, it has been shown that connected components of $G$ without captive buyers have zero revenue in every exact equilibrium and can be assigned price zero~\citep[Theorem~4.5]{BLN13}. Consequently, our structural analysis in Sections~\ref{sec:finite}--\ref{sec:exact-algorithm} focuses on connected components containing at least one captive seller, where every seller earns strictly positive equilibrium revenue.


To understand the functional form of exact equilibria on components with captive buyers, recall from~\cite{BLN13} that on any open price interval where a fixed subset of sellers is actively randomizing, every active seller $i$ must earn their constant equilibrium revenue $u_i$ at all prices $p$ in that interval:
\begin{align*}
 p\rbr{\alpha_i + \sum_{j:\cbr{i,j}\in E}\beta_{ij}\rbr{1 - F_j(p)}} = u_i.
\end{align*}
Dividing both sides by $p$ shows that the linear combination of the neighbors' CDFs must equal $u_i/p - \alpha_i$, which is an affine function of the reciprocal price $1/p$. Based on this indifference condition,~\cite{BLN13} showed that on any interval where the set of active sellers does not change, the sellers' equilibrium CDFs can be written as affine functions of $1/p$ (and must take this form whenever the active adjacency submatrix is nonsingular).

Motivated by this characterization, we say that a mixed strategy profile $\bF$ is \emph{reciprocal-affine} if there exists a finite partition
\begin{align*}
 0 = p_0 < p_1 < \cdots < p_K = 1
\end{align*}
of $(0,1)$ into $K$ intervals such that, on each open interval $(p_{k-1}, p_k)$ ($1 \leq k \leq K$), every seller $i$'s CDF takes the form
\begin{align*}
 F_i(p) = a_{ik} + \frac{b_{ik}}{p}
\end{align*}
for some constants $a_{ik}, b_{ik} \in \R$ (supplemented by possible point masses at $p = 1$, and $P_i = 0$ on components without captive buyers). Because the $2n$ coefficients $\cbr{a_{ik}, b_{ik}}_{i \in [n]}$ completely specify all sellers' distributions on each interval $(p_{k-1}, p_k)$, the size of a finite description of $\bF$ is governed by the number of intervals $K$. Accordingly, we define the \emph{structural complexity} of a reciprocal-affine profile $\bF$ as the minimum number of intervals $K$ in a common partition of $(0,1)$ on which $\bF$ is reciprocal-affine. Similarly, the \emph{structural complexity of an achievable revenue vector $\bu$} is the minimum structural complexity among all exact equilibria $\bF$ that realize $\bu$.

The support of $F_i$, denoted $\operatorname{supp}(F_i)$, is the smallest closed subset of $[0,1]$ carrying probability one, and its \emph{support boundaries} are the topological boundary points of $\operatorname{supp}(F_i)$ in $[0,1]$. Notice that on each open interval $(p_{k-1}, p_k)$ of a reciprocal-affine partition, each seller's CDF $F_i(p) = a_{ik} + b_{ik}/p$ is smooth and is either strictly increasing (when $b_{ik} < 0$) or constant (when $b_{ik} = 0$). Thus, no point inside $(p_{k-1}, p_k)$ can be a support boundary point of $F_i$. Every support boundary point of any seller in $(0,1)$ must therefore belong to the set of partition breakpoints $\cbr{p_1,\ldots,p_{K-1}}$. In particular, if the union of the sellers' support boundaries in $(0,1)$ contains $B$ distinct prices, then any common reciprocal-affine partition requires at least $K \geq B + 1$ intervals.


\paragraph{Computational model}
For algorithmic problems, we assume the standard bit-complexity model in which all numerical inputs ($\alpha_i, \beta_{ij}$, and $\eps$) are rational numbers with total binary encoding length $\cL$. In our exact algorithm (Section~\ref{sec:exact-algorithm}), where the exact equilibrium revenues and reciprocal-price breakpoints may be irrational real algebraic numbers, each algebraic output quantity is represented in standard form by an integer polynomial of which it is a root together with an isolating rational interval that uniquely identifies that root.

\section{$\PPAD$-completeness}
\label{sec:bh-hardness}

To prove Theorem~\ref{thm:bh-hardness}, we proceed in two stages. First,
we reduce the $\PPAD$-hard binary polymatrix game to an intermediate
pricing game in which each seller is restricted to a two-price menu and
receives linear unary payoffs. Second, we replace these artificial menus
and unary payoffs with unconstrained continuous pricing in $[0,1]$ by
attaching price-anchor gadgets to every seller, showing that every
$\eps_0$-equilibrium of the resulting Bertrand network can be decoded
back to an approximate equilibrium of the source game. We conclude the
section by formalizing the finite grid search relation and establishing
$\PPAD$ membership.

\subsection{A Binary Simulation with $12$ Replicas}
\label{sec:bh-binary}

Our reduction starts from the $\PPAD$-hard problem of computing a
$1/20$-Nash equilibrium in a bipartite, degree-three, two-action
polymatrix game with total payoffs in $[0,1]$, as established
by~\cite[Theorem~4.4]{DFHM24}. For each source player $i$, let
$D_i^{\mathrm{orig}}(\bx)=c_i^{\mathrm{orig}}+\sum_{j\in N(i)}d_{ij}^{\mathrm{orig}}x_j$
denote its expected payoff difference between action one and action zero
when its neighbors play action one with probabilities $\bx$. We fix a
rounding precision $\eta>0$ and round each coefficient
$c_i^{\mathrm{orig}}, d_{ij}^{\mathrm{orig}}$ toward zero to the nearest
integer multiple of $\eta$, obtaining the rounded payoff difference
\begin{align}
 D_i(\bx)=c_i+\sum_{j\in N(i)}d_{ij}x_j.
 \label{eq:bh-source}
\end{align}
Because $D_i^{\mathrm{orig}}(\bx)$ maps $[0,1]^{N(i)}$ into $[-1,1]$,
evaluating at $\bx=\mathbf{0}$ gives $\abs{c_i^{\mathrm{orig}}}\leq 1$,
and taking the difference between the maximum and minimum values over
the hypercube $[0,1]^{N(i)}$ gives
$\sum_{j\in N(i)}\abs{d_{ij}^{\mathrm{orig}}}\leq 2$. Rounding toward
zero preserves these bounds, so that $\abs{c_i}\leq 1\leq 3$,
$\abs{d_{ij}}\leq 2$, and $\sum_{j\in N(i)}\abs{d_{ij}}\leq 2$. Since each
source player has degree at most three in the polymatrix game, the
rounded payoff difference satisfies
\begin{align*}
 \sup_{\bx\in[0,1]^{N(i)}}\abs{D_i^{\mathrm{orig}}(\bx)-D_i(\bx)}\leq 4\eta.
\end{align*}

In the intermediate binary-menu game, each seller is assigned one of four
two-price menus $\cbr{a_i,b_i}$ corresponding to roles $L,M,H,Z$, along
with unary payoffs $(d_0,d_1)$ for playing low or high:
\begin{align*}
\begin{array}{c|cccc}
 \text{role}&L&M&H&Z\\ \hline
 \text{low price } a&1/5&1/4&1/3&2/5\\
 \text{high price } b&3/5&2/3&3/4&4/5
\end{array}
\end{align*}
Note that all eight menu prices are distinct with a minimum separation of
$1/20$, and every low price is strictly smaller than every high price.
Let $x_i$ denote the probability that seller $i$ chooses its high price
$b_i$. Whenever two adjacent sellers $i$ and $j$ have interleaved menus
satisfying $a_i<a_j<b_i<b_j$, seller $i$ always wins a shared market of
mass $\beta$ at price $a_i$ and wins at price $b_i$ only when seller $j$
plays $b_j$, whereas seller $j$ wins at $a_j$ only when seller $i$ plays
$b_i$ and never wins at $b_j$. Consequently, the shared market
contributes
\begin{align}
 \Delta_i=\beta b_i x_j-\beta a_i,\quad
 \Delta_j=-\beta a_j x_i
 \label{eq:bh-cross}
\end{align}
to the respective high-minus-low payoff differences of sellers $i$ and $j$.

\paragraph{The replica path}
For each source player, we create $12$ replicas consisting of three
copies of each role-sign pair $L^+,H^+,L^-,H^-$. If $x$ denotes a
replica's actual probability of playing its high price and $\Delta$
denotes its actual high-minus-low payoff difference, we define a positive
replica to have abstract probability $t=x$ and abstract payoff difference
$g=\Delta$, and a negative replica to have abstract probability $t=1-x$
and abstract payoff difference $g=-\Delta$. We assign weights $w_L=1$ to
all role-$L$ replicas and $w_H=24/5$ to all role-$H$ replicas, so that
the total weight across the $12$ replicas is
\begin{align*}
 w_L=1,\quad w_H=24/5,\quad W=\sum_v w_v=174/5.
\end{align*}
We then arrange the six positive replicas and six negative replicas along
a path in the order (see Figure~\ref{fig:replicas})
\begin{align*}
 H^+_1,\ L^+_1,\ H^+_2,\ L^+_2,\ H^+_3,\ L^+_3
 \quad\text{and}\quad
 L^-_1,\ H^-_1,\ L^-_2,\ H^-_2,\ L^-_3,\ H^-_3.
\end{align*}
Consecutive replicas of the same sign are connected by an equality link
passing through a dedicated role-$M$ auxiliary seller, with edge mass
$4M_0$ incident to the $L$-replica and $3M_0/2$ incident to the
$H$-replica. In addition, $L^+_3$ and $L^-_1$ are connected by a
complementation link passing through a role-$Z$ auxiliary seller with
edge mass $5M_0/2$ on both incident edges. This construction introduces
$12$ replicas and $11$ auxiliary sellers for each source player.

To center the internal link payoffs, each equality link adds unary
high-minus-low payoff offsets of $-2M_0/5$ at its $L$-replica, $M_0/4$ at
its $H$-replica, and $3M_0/8$ at its $M$-auxiliary, while the
complementation link adds unary offsets of $-M_0/4$ at both endpoint
replicas and $M_0$ at its $Z$-auxiliary. Whenever a replica participates
in two internal links, its unary offsets are summed.

We orient each equality link $e=(v,w)$ from its $L$-replica $v$ to its
$H$-replica $w$, and define its auxiliary variable $y_e$ to be the
$M$-seller's high-price probability on a positive link and one minus
that probability on a negative link. Similarly, we orient the
complementation link from $v=L^+_3$ to $w=L^-_1$ and set $y_e$ to be the
$Z$-seller's high-price probability. Substituting these parameters
into~\eqref{eq:bh-cross} shows that on every oriented internal link
$e=(v,w)$,
\begin{align}
 \begin{aligned}
 \text{auxiliary abstract difference}&=M_0(t_w-t_v),\\
 w_v\,\text{signal at }v&=k_e(y_e-\tfrac12),\\
 w_w\,\text{signal at }w&=-k_e(y_e-\tfrac12),
 \end{aligned}
 \quad\text{where}\quad
 \frac{k_e}{M_0}=
 \begin{cases}
 12/5&\text{on an equality link},\\
 3/2&\text{on the complement link}.
 \end{cases}
 \label{eq:bh-canonical}
\end{align}
In particular, the weighted signals sent to the two endpoints of every
internal link have equal magnitude and opposite signs.

Recall that in an unnormalized $\delta$-equilibrium of the binary-menu
game, any player with abstract probability $t$ and abstract payoff
difference $g$ satisfies $(a-t)g\leq\delta$ for both $a\in\cbr{0,1}$. 
Let $E_v$ denote the external contribution to replica $v$'s abstract payoff difference coming from edges and unary payoff offsets outside its replica path, and suppose that $\sum_v w_v\abs{E_v}\leq G$ for a uniform constant $G>0$.
In the
following lemma, we show that choosing the internal link scale
$M_0 \geq 8G$ forces all $12$ replica probabilities to lie within
$O(\delta/M_0)$ of one another and allows us to aggregate their external
payoff differences while canceling the internal consensus signals up to
$O(\delta)$.

\begin{lemma}[Consensus and aggregation]
\label{lem:bh-consensus}
Suppose $\sum_v w_v\abs{E_v}\leq G$ and $M_0\geq 8G$. Then in every
unnormalized $\delta$-equilibrium of the binary-menu game, the $12$
replicas of a source player satisfy
\begin{align}
 \max_v t_v-\min_v t_v&\leq\frac{88W}{M_0}\delta,
 \label{eq:bh-diameter}\\
 \sum_v(a-t_v)w_v E_v&\leq\frac{603}{10}\delta
 \quad\text{for } a\in\cbr{0,1}.
 \label{eq:bh-aggregate}
\end{align}
\end{lemma}
\begin{proof}
Assume first that $\delta>0$ and set $\tau=8W\delta/M_0$. Suppose for the
sake of contradiction that the diameter of the $12$ probabilities $t_v$
exceeds $11\tau$. Then two consecutive values of $t_v$ in sorted order
must be separated by a gap greater than $\tau$. Let $S$ denote the
nonempty subset of replicas above this gap, so that every $v\in S$
satisfies $t_v>\tau$.

For any internal link $e=(v,w)$ crossing the cut between $S$ and its
complement, the auxiliary seller's abstract payoff difference has
magnitude $M_0\abs{t_w-t_v}>M_0\tau$, which implies that its suboptimal
action receives probability at most $\delta/(M_0\tau)=1/(8W)\leq 1/4$.
By~\eqref{eq:bh-canonical}, the weighted signal exerted by link $e$ on
its endpoint in $S$ is therefore at most $-k_e/4\leq -3M_0/8$. Because
the weighted signals on links with both endpoints in $S$ cancel in pairs
and the replica path is connected, summing $w_v g_v$ over all $v\in S$
gives
\begin{align*}
 \sum_{v\in S}w_v g_v
 \leq \sum_{v\in S}w_v E_v-\frac{3M_0}{8}
 \leq G-\frac{3M_0}{8}\leq-\frac{M_0}{4}.
\end{align*}
On the other hand, since each replica $v\in S$ can deviate to abstract
action $a=0$, we have $-t_v g_v\leq\delta$, and hence
$g_v\geq -\delta/t_v > -\delta/\tau$. Summing over $v\in S$ yields
\begin{align*}
 \sum_{v\in S}w_v g_v > -\frac{W\delta}{\tau}=-\frac{M_0}{8},
\end{align*}
which contradicts the upper bound above. When $11\tau\geq 1$, the
bound~\eqref{eq:bh-diameter} holds trivially, and the case $\delta=0$
follows by taking the limit $\delta\downarrow 0$.

To establish~\eqref{eq:bh-aggregate}, we sum the weighted deviation
inequalities $(a-t_v)w_v g_v\leq w_v\delta$ over all $12$ replicas $v$.
By~\eqref{eq:bh-canonical}, the internal link contributions sum to
\begin{align*}
 \sum_{e=(v,w)}k_e(t_w-t_v)(y_e-\tfrac12).
\end{align*}
For each link $e=(v,w)$, the auxiliary seller can deviate to the uniform
mixture $1/2$, which gives $(\tfrac12-y_e)M_0(t_w-t_v)\leq\delta$, or
equivalently $M_0(t_w-t_v)(y_e-\tfrac12)\geq -\delta$. Subtracting these
internal contributions across the ten equality links and one
complementation link therefore yields
\begin{align*}
 \sum_v(a-t_v)w_v E_v
 \leq\rbr{W+10\cdot\frac{12}{5}+\frac32}\delta
 =\frac{603}{10}\delta,
\end{align*}
as desired.
\end{proof}

\paragraph{Simulating the source interactions}
Identifying the replica signs with $\pm 1$, an external edge of mass
$\beta$ between $L_i^s$ and $H_j^t$ contributes the weighted coefficient
pair $st\beta(3/5,-8/5)$ to the abstract payoff differences of source
players $i$ and $j$, while the opposite orientation $H_i^s$--$L_j^t$
contributes $st\beta(-8/5,3/5)$. Because the vectors $(3/5,-8/5)$ and
$(-8/5,3/5)$ are linearly independent, for each source edge $\cbr{i,j}$
with $i<j$ we can define
\begin{align}
 z_1=-\frac{3d_{ij}+8d_{ji}}{11},\quad
 z_2=-\frac{8d_{ij}+3d_{ji}}{11},
 \label{eq:bh-decomposition}
\end{align}
and place an edge of mass $\abs{z_1}$ between
$L_i^{\operatorname{sign}(z_1)}$ and $H_j^+$ together with an edge of
mass $\abs{z_2}$ between $H_i^{\operatorname{sign}(z_2)}$ and $L_j^+$
(omitting edges with zero mass). Since $\abs{d_{ij}},\abs{d_{ji}}\leq 2$,
both edge masses are at most $2$. Moreover, because each source player has
degree at most three, we can assign its incident external edges to
distinct replicas within each role-sign pair.

To cancel the constant terms introduced by~\eqref{eq:bh-cross} on an
external edge of mass $\beta$ between signs $s_L$ and $s_H$, we add the
unary high-minus-low payoff offsets
\begin{align}
 h_L=\beta\rbr{a_L-b_L\frac{1-s_H}{2}},\quad
 h_H=\beta a_H\frac{1-s_L}{2}
 \label{eq:bh-intercepts}
\end{align}
at its $L$ and $H$ endpoints. Substituting $x=(1-s)/2+st$ confirms that
only the desired bilinear interaction terms remain. Finally, we add the
rounded source constant $c_i$ to the unary difference of $L^+_{i,1}$ and
shift each seller's unary payoffs so that its low-action unary payoff is
zero.

Because each source player has at most three neighbors and each external
edge has mass at most $2$, the weighted sum of external coefficient
magnitudes for player $i$ is bounded by
\begin{align*}
 L_*=3\rbr{\frac35\cdot 2+\frac85\cdot 2}=\frac{66}{5},
\end{align*}
so that $G=3+L_*=\frac{81}{5}.$
If we define $t_i=t_{L^+_{i,1}}$ as the decoded action-one probability for
source player $i$, then replacing each replica probability $t_v$
in~\eqref{eq:bh-aggregate} by $t_i$ and each neighbor replica probability
by $t_j$ incurs an additional error of at most $(G+L_*)$ times the
diameter bound~\eqref{eq:bh-diameter}. Consequently, for both
$a\in\cbr{0,1}$, we obtain
\begin{align}
 (a-t_i)D_i(\bt)\leq \cC(M_0)\delta,
 \quad\text{where}\; \; 
 \cC(M_0)=\frac{603}{10}+(G+L_*)\frac{88W}{M_0}.
 \label{eq:bh-source-decoding}
\end{align}
By construction, every replica has at most two internal consensus edges
and one external source edge, so its total incident strategic mass is at
most $8M_0+2$, and its total unary payoff magnitude is at most
$4M_0/5+4/5+3$. Thus, any constant $E\geq 8M_0+4$ upper-bounds both the
strategic incident mass and the unary payoff magnitude of every seller in
the binary-menu game.

\subsection{A Robust Interior-price Anchor}
\label{sec:bh-anchor}

To eliminate the artificial two-price menus, we next construct a
three-seller path gadget $R$--$S$--$T$ whose terminal seller $T$
concentrates its price distribution near a target price $q\in[1/5,4/5]$.
In this gadget, seller $R$ has captive mass $\alpha_R=\frac{q}{1-q}B$,
sellers $S$ and $T$ have no captive buyers, the internal edges have
masses $\beta_{RS}=B$ and $\beta_{ST}=C$, and only $T$ connects to the
rest of the graph via external edges of total mass $D$. In the next
lemma, we prove that when the market scales satisfy $B \gg C \gg D$, any
$\eps$-equilibrium forces $\Ex{\abs{P_T-q}}$ to be small.

\begin{lemma}[Price anchor]
\label{lem:bh-anchor}
Fix $q\in[1/5,4/5]$ and consider the three-seller path $R$--$S$--$T$ with
$\alpha_R=\frac{q}{1-q}B$, $\beta_{RS}=B$, $\beta_{ST}=C$, and external
mass $D$ incident to $T$. If $C/B\leq 1$, $D/C\leq 1$, and
$5\eps+2C/B\leq 10^{-6}$, then every $\eps$-equilibrium satisfies
\begin{align}
 \Ex{\abs{P_T-q}}
 \leq K\rbr{\sqrt{\eps+C/B}+\eps+D/C},
 \label{eq:bh-anchor}
\end{align}
where $K = 1000$.
\end{lemma}
\begin{proof}
Set $a=q/(1-q)\in[1/4,4]$ and $c=a+1=1/(1-q)\in[5/4,5]$. If we remove
edge $ST$ and normalize the remaining $RS$ duopoly by $B$, seller $R$ has
accessible mass $cB\leq 5B$ and seller $S$ has accessible mass
$B+C\leq 2B$. Since edge $ST$ alters $S$'s payoff by at most $C$, both $R$
and $S$ can increase their expected revenue by at most
$\delta_0=5\eps+2C/B\leq 10^{-6}$ by deviating in the unit-scaled $RS$
duopoly. Similarly, removing $T$'s external edges and dividing by $C$
shows that $T$ can increase its expected revenue per unit of mass on $ST$
by at most $2\eps+D/C$ against $S$'s price distribution.

In the unit-scaled $RS$ duopoly, let $U$ and $V$ denote the actual
expected revenues of $R$ and $S$, and let $P$ and $Q$ denote their
supremal pure-deviation revenues. Writing $Y(p)=\Pr{P_R\geq p}$ and
$Z(p)=\Pr{P_S\geq p}$ for their left-continuous survival functions,
undercutting any price $p\in(0,1]$ yields
\begin{align*}
 p(a+Z(p))\leq P,\quad pY(p)\leq Q,
\end{align*}
with $a\leq P\leq c$ and $0\leq Q\leq 1$. Because the shared market on
$RS$ is won by the minimum price $\min(P_R,P_S)$ regardless of ties, the
joint revenue of $R$ and $S$ is given by
\begin{align*}
 U+V=\int_0^1 Y(p)(a+Z(p))\,dp,
\end{align*}
and satisfies $P+Q-(U+V)\leq 2\delta_0.$
Defining the upper envelopes $\overline{Y}(p)=\min(1,Q/p)$ and
$\overline{Z}(p)=\min(1,P/p-a)$, we have $Y(p)\leq\overline{Y}(p)$ and
$Z(p)\leq\overline{Z}(p)$ for all $p\in[0,1]$. Direct integration gives
\begin{align*}
 I(P,Q)\defeq\int_0^1\overline{Y}(p)(a+\overline{Z}(p))\,dp
 =
 \begin{cases}
 2cQ-PQ+cQ\log(P/(cQ)),&Q\leq P/c,\\
 2P-PQ+P\log(cQ/P),&Q\geq P/c.
 \end{cases}
\end{align*}
For any fixed $P\in[a,c]$, differentiating $D_0(P,Q)\defeq P+Q-I(P,Q)$
with respect to $Q$ gives $\partial_Q D_0(P,Q)=P-a-c\log(P/(cQ))$ for
$Q\leq P/c$ and $\partial_Q D_0(P,Q)=1+P-P/Q$ for $Q\geq P/c$. Hence,
$D_0(P,Q)$ is uniquely minimized at $Q_*=(P/c)e^{-(P-a)/c}$, where
\begin{align*}
 D_0(P,Q)\geq D_0(P,Q_*)=P\rbr{1-e^{-(P-a)/c}}\geq 0.
\end{align*}
Since $0\leq P-a\leq 1$ and $c/a=1/q\leq 5$, the inequality
$D_0(P,Q)\leq P+Q-(U+V)\leq 2\delta_0$ implies that
\begin{align*}
 0\leq P-a\leq 10e\delta_0<28\delta_0.
\end{align*}
Furthermore, $\partial_Q^2 D_0(P,Q)$ equals $c/Q$ for $Q<P/c$ and
$P/Q^2$ for $Q>P/c$, both of which are at least $1/4$. A second-order
Taylor expansion around $Q_*$ therefore yields
$\abs{Q-Q_*}\leq 4\sqrt{\delta_0}$. Since $Q_*$ is $1$-Lipschitz in $P$
and equals $a/c=q$ at $P=a$, it follows that
$\abs{Q-q}\leq 4\sqrt{\delta_0}+28\delta_0$ and $Q\geq 1/10$.

Next, because $\overline{Y}(p)\geq Q$ and $a+Z(p)\geq 0$, expanding
$(\overline{Y}-Y)(a+Z)+\overline{Y}(\overline{Z}-Z)$ gives
\begin{align*}
 Q\int_0^1(\overline{Z}(p)-Z(p))\,dp
 \leq I(P,Q)-(U+V)\leq 2\delta_0.
\end{align*}
Since $\overline{Z}(p)=1$ on $[0,q]$ and $Z$ is nonincreasing, setting
$\gamma=\sqrt{2\delta_0/Q}<q$ ensures that
$Z(q-\gamma)\geq 1-2\delta_0/(Q\gamma)$. By undercutting $q-\gamma$,
seller $T$ can achieve expected revenue at least
$(q-\gamma)Z(q-\gamma)\geq q-2\sqrt{2\delta_0/Q}\geq q-9\sqrt{\delta_0}$
per unit of mass on $ST$. Conversely, any price $p\in[0,1]$ played by $T$
earns at most
\begin{align*}
 pZ(p)\leq\min(p,P-ap)\leq q-\min(1,a)\abs{p-q}+(P-a).
\end{align*}
Since $\min(1,a)\geq 1/4$ and $T$'s maximum deviation gain per unit of
mass on $ST$ is at most $2\eps+D/C$, taking expectations over $P_T$
yields
\begin{align*}
 \Ex{\abs{P_T-q}}
 \leq 4\rbr{28\delta_0+9\sqrt{\delta_0}+2\eps+D/C}
 \leq 148\sqrt{\delta_0}+4(2\eps+D/C),
\end{align*}
which establishes~\eqref{eq:bh-anchor}.
\end{proof}

\subsection{Binary Actions via Continuous Prices}
\label{sec:bh-menu}

We now attach two price anchors to each computational seller $i$ in the
binary-menu game to enforce its menu $\cbr{a_i,b_i}$ and unary payoffs
$(d_{i,0},d_{i,1})$ under unrestricted pricing in $[0,1]$. Let $E\geq 1$
be the uniform upper bound on each seller's strategic incident mass and
unary payoff magnitude from Section~\ref{sec:bh-binary}. Given a baseline
parameter $U\geq 4E$, a band half-width $\sigma\in(0,1/100]$, a deviation
offset $r\in(0,\sigma/2]$, and an anchor accuracy $\rho\in(0,r^2]$, we
connect seller $i$ to two anchor terminals $T_{a_i}$ and $T_{b_i}$ using
edge masses
\begin{align}
 W_1=\frac{U+d_{i,1}}{b_i},\quad
 W_0=\frac{U+d_{i,0}}{a_i}-W_1.
 \label{eq:bh-menu-masses}
\end{align}
When each computational seller is in an unnormalized $\zeta$-equilibrium,
we bound the probability of playing outside the $\sigma$-neighborhoods
of $a_i$ and $b_i$ by
\begin{align}
 \chi=\frac{6E+64Ur+2\zeta}{U\sigma},
 \label{eq:bh-menu-error}
\end{align}
and set
\begin{align*}
    \Delta_{\mathrm{menu}}=\zeta+32Ur+2E\sigma+5E\chi.
\end{align*}
In the next lemma, we show that thresholding each seller's continuous
price at the menu midpoint $(a_i+b_i)/2$ yields an unnormalized
$\Delta_{\mathrm{menu}}$-equilibrium of the intermediate binary-menu
game.

\begin{lemma}[Menu implementation]
\label{lem:bh-menu}
Suppose every computational seller $i$ has no captive buyers, faces two
anchor terminals satisfying $\Ex{\abs{P_{T_q}-q}}\leq\rho$ for
$q\in\cbr{a_i,b_i}$ with attachment masses~\eqref{eq:bh-menu-masses}, and
can increase its expected revenue by at most $\zeta$ via any unilateral
deviation in $[0,1]$. Then the decoded binary probabilities
$x_i=\Pr{P_i>(a_i+b_i)/2}$ form an unnormalized
$\Delta_{\mathrm{menu}}$-equilibrium of the restricted binary-menu game,
where $\chi$ and $\Delta_{\mathrm{menu}}$ are given
by~\eqref{eq:bh-menu-error}.
\end{lemma}
\begin{proof}
Since $b_i\geq 2a_i$ for all four menu roles and $U\geq 4E$, both $W_1$
and $W_0\geq(U-3E)/b_i$ are strictly positive, and their sum satisfies
$W_0+W_1=(U+d_{i,0})/a_i\leq 6.25U$. Let $B_i(p)$ denote seller $i$'s
expected revenue from its two anchor edges at price $p$. By Markov's
inequality and the condition $\rho\leq r^2$, we have
$\Pr{P_{T_q}\geq p}\leq\Ind{p\leq q+r}+r$, which yields the upper
envelope
\begin{align}
 B_i(p)\leq p\rbr{W_0\Ind{p\leq a_i+r}+W_1\Ind{p\leq b_i+r}}+r(W_0+W_1).
 \label{eq:bh-envelope}
\end{align}
We refer to the closed intervals $[a_i-\sigma,a_i+\sigma]$ and
$[b_i-\sigma,b_i+\sigma]$ as good bands $0$ and $1$, and to all other
prices in $[0,1]$ as bad prices. Because $(a_i+\sigma)/b_i\leq 3/5$,
$(3/5)(U+E)\leq U-E\leq U+d_{i,0}$, and $\min(W_0+W_1,W_1)\geq U/2$,
evaluating~\eqref{eq:bh-envelope} on each region gives
\begin{align*}
 B_i(p)&\leq U+E+20Ur &&\text{for all } p\in[0,1],\\
 B_i(p)&\leq U+d_{i,k}+20Ur &&\text{for } p \text{ in good band } k\in\cbr{0,1},\\
 B_i(p)&\leq U+E-U\sigma/2+20Ur &&\text{for all bad prices } p.
\end{align*}
Conversely, if seller $i$ deviates to $q_k-r$ for $q_0=a_i$ or $q_1=b_i$,
Markov's inequality ensures that $P_{T_{q_k}}>q_k-r$ with probability at
least $1-r$, and therefore
\begin{align*}
 B_i(q_k-r)\geq U+d_{i,k}-12Ur.
\end{align*}

Let $\chi_i$ denote the probability that $P_i$ falls in the bad region.
Since seller $i$'s strategic revenue from non-anchor edges lies in
$[0,E]$, comparing its actual expected revenue against the deviation
$a_i-r$ and using the $\zeta$ deviation bound implies that
\begin{align*}
 \frac{\chi_i U\sigma}{2}\leq 3E+32Ur+\zeta,
 \quad\text{so}\quad
 \chi_i\leq\chi.
\end{align*}
Furthermore, writing $A_i(p)=\Ind{p>(a_i+b_i)/2}$, the regional bounds
on $B_i(p)$ combine into the pointwise inequality
\begin{align*}
 B_i(p)\leq U+d_{i,A_i(p)}+20Ur+2E\Ind{p\text{ is bad}}.
\end{align*}
Notice that the large baseline $U$ cancels when we compare $B_i(P_i)$
against $B_i(q_k-r)$, which means that bad prices incur an anchor penalty
of at most $2E\chi_i$.

On any strategic edge $\cbr{i,j}$, all menu prices are separated by at
least $1/20>2\sigma$. Consequently, whenever $P_i$ and $P_j$ both lie in
their respective good bands, their relative order matches the order of
their decoded menu prices, and their expected edge revenue differs from
the discrete menu revenue by at most $\beta_{ij}\sigma$. Accounting for
the probability of bad prices at $i$ and its neighbors $j$, seller $i$'s
actual strategic revenue exceeds its decoded binary revenue by at most
$E\sigma+E\chi_i+\sum_j\beta_{ij}\chi_j$, while deviating to $q_k-r$
achieves at least the pure binary deviation revenue minus
$Er+\sum_j\beta_{ij}\chi_j$. Combining these anchor and strategic bounds
with seller $i$'s $\zeta$ deviation bound shows that the decoded binary
profile has maximum deviation gain at most
\begin{align*}
 \zeta+32Ur+E\sigma+Er+3E\chi_i+2\sum_j\beta_{ij}\chi_j
 \leq\Delta_{\mathrm{menu}}.
\end{align*}
\end{proof}

\subsection{Finishing the Proof}
\label{sec:bh-completion}

We now combine the binary-menu simulation of Section~\ref{sec:bh-binary}
with the price-anchor gadgets of Sections~\ref{sec:bh-anchor}
and~\ref{sec:bh-menu} to establish the $\PPAD$-hardness part of
Theorem~\ref{thm:bh-hardness}.

\begin{proof}[Proof of Theorem~\ref{thm:bh-hardness} ($\PPAD$-hardness)]
Starting from the rounded source polymatrix game, we construct the
replica and auxiliary graph of Section~\ref{sec:bh-binary} and attach two
disjoint three-seller price anchors to each computational seller. For
each anchor attachment of mass $D=W_k$, we set the internal anchor masses
to
\begin{align}
 C=\frac{10KD}{\rho},\quad
 B=\frac{100K^2C}{\rho^2},\quad
 \alpha_R=\frac{q}{1-q}B,\quad\text{where } K=1000.
 \label{eq:bh-anchor-scales}
\end{align}
We choose the universal constants $(M_0,E,U,\sigma,r,\rho,\zeta,\eps_0,\eta,\gamma)$
according to the explicit assignment in Appendix~\ref{sec:bh-parameters},
which satisfies
\begin{align}
 \eps_0\leq\min\cbr{\frac{\zeta}{11U},\frac{\rho^2}{100K^2}}
 \quad\text{and}\quad
 \cC(M_0)\Delta_{\mathrm{menu}}+4\eta+3\gamma<\frac{1}{20},
 \label{eq:bh-final-obligations}
\end{align}
where $\gamma$ denotes the additive error tolerance of the CDF evaluation
oracle.

Suppose we are given any $\eps_0$-equilibrium of the constructed Bertrand
network. Because each computational seller has total accessible mass at
most $E+10U\leq 11U$, the first condition in~\eqref{eq:bh-final-obligations}
ensures that every computational seller can increase its expected revenue
by at most $\eps_0(11U)\leq\zeta$ via any unilateral deviation in
$[0,1]$. Moreover, for every price anchor,~\eqref{eq:bh-anchor-scales}
gives $D/C=\rho/(10K)$ and $C/B=\rho^2/(100K^2)$, so
Lemma~\ref{lem:bh-anchor} implies that its terminal seller $T$ satisfies
\begin{align*}
 \Ex{\abs{P_T-q}}
 \leq K\rbr{\sqrt{\eps_0+C/B}+\eps_0+D/C}
 \leq\rbr{\frac{\sqrt{2}}{10}+\frac{1}{100K}+\frac{1}{10}}\rho<\rho.
\end{align*}
Therefore, all hypotheses of Lemma~\ref{lem:bh-menu} hold, and the
decoded binary probabilities $x_i=\Pr{P_i>(a_i+b_i)/2}$ form an
unnormalized $\Delta_{\mathrm{menu}}$-equilibrium of the intermediate
binary-menu game. Applying~\eqref{eq:bh-source-decoding} from
Lemma~\ref{lem:bh-consensus} then bounds each source player's maximum
deviation gain in the rounded polymatrix game by
$\cC(M_0)\Delta_{\mathrm{menu}}$. Adding at most $4\eta$ from rounding
the source coefficients and at most $3\gamma$ from evaluating
$1-F_i((a_i+b_i)/2)$ within additive error $\gamma$, the second condition
in~\eqref{eq:bh-final-obligations} guarantees that the decoded strategy
profile is a $1/20$-Nash equilibrium of the original polymatrix game.

Finally, we verify the structural restrictions on the constructed
network. Each source player is represented by $23$ computational sellers,
and attaching two three-seller anchors to each yields $161$ sellers per
source player. Every replica has degree at most five (at most two
internal path links, one external source edge, and two anchor edges),
auxiliary sellers have degree four, and anchor sellers have degree at
most two, with captive markets appearing only at the degree-one anchor
leaves $R$. Coloring all replicas with the bipartition color of their
source player, coloring all $M$ and $Z$ auxiliaries with the opposite
color, and alternating colors along each anchor path confirms that the
constructed graph is bipartite. Lastly, rescaling all market masses by
the fixed integer $Q_0$ in Appendix~\ref{sec:bh-parameters} places all
positive market masses in a fixed finite set of integers without altering
the $\eps_0$-equilibrium condition.
\end{proof}

\subsection{PPAD Membership}
\label{sec:bh-membership}

To complete the proof of Theorem~\ref{thm:bh-hardness}, we formalize the
finite price-grid search problem and show that it belongs to $\PPAD$. For
a given accuracy $\eps>0$, let $J=\Cel{2/\eps}$ and $h=1/J$, and consider
price distributions supported on the uniform grid $\cbr{0,h,2h,\ldots,1}$.
We define the search problem as finding a profile of rational probability
distributions supported on $\cbr{0,h,2h,\ldots,1}$ of polynomially
bounded bit length that forms an $\eps$-equilibrium of the continuous
Bertrand network against all real-price deviations $p\in[0,1]$.

\begin{proof}[Proof of Theorem~\ref{thm:bh-hardness} ($\PPAD$ membership)]
Fix $\eps>0$, set $J=\Cel{2/\eps}$ and $h=1/J$, and restrict all sellers'
pure strategies to the finite price grid $\cbr{0,h,2h,\ldots,1}$. For any
continuous price deviation $p\in[0,1]$ against a grid-supported strategy
profile $\bF_{-i}$, let $p^-=h\Flr{p/h}$ be the nearest grid point at or
below $p$. Rounding the deviation price from $p$ down to $p^-$ preserves
every shared market won at $p$, since any opponent charging strictly more
than $p$ also charges strictly more than $p^-$. Because rounding down
reduces the price received on those won markets by at most $h$, we have
\begin{align}
 u_i(p,\bF_{-i})\leq u_i(p^-,\bF_{-i})+hM_i.
 \label{eq:bh-grid}
\end{align}

Observe that the restricted finite game on the grid $\cbr{0,h,2h,\ldots,1}$
is a polymatrix game with captive-market payoffs $p_i\alpha_i$, which can
be placed into standard polymatrix form by introducing a single-action
dummy player connected to each seller $i$. It is shown
by~\cite[Corollary~25(3)]{EY10} that computing an exact Nash equilibrium
of a finite polymatrix game belongs to $\PPAD$. Moreover,
by~\eqref{eq:bh-grid}, any exact equilibrium of the restricted grid game
is an $(hM_i)$-unnormalized equilibrium, and hence an
$(\eps/2)$-equilibrium of the continuous Bertrand network against all
deviations $p\in[0,1]$.

Furthermore, once the support of an exact grid equilibrium is fixed, the
support, indifference, and best-response conditions form a system of
linear equations and inequalities with rational coefficients. By
Carath\'eodory's theorem~\citep{TijsBorwein1980} and Cramer's
rule~\citep{Mizunoetal2015,EisenbrandVempala2016} (see
also~\cite[Corollary~25(3)]{EY10}), this linear system admits a basic
feasible solution of polynomial bit length that remains an exact grid
equilibrium. Finally, given any candidate grid distribution profile,
one can verify the continuous $\eps$-equilibrium
condition~\eqref{eq:model-equilibrium} in polynomial time: because each
seller's expected revenue $u_i(p,\bF_{-i})$ is affine on every open
interval between consecutive grid points, its supremum over $p\in[0,1]$
is attained among the grid points, their left-hand limits, and price one.
Setting $\eps=\eps_0$ establishes that the search problem belongs to
$\PPAD$, completing the proof of Theorem~\ref{thm:bh-hardness}.
\end{proof}

\section{EPTAS Algorithm}
\label{sec:eptas}

When both the maximum degree $\Delta$ of the graph and the number of
captive sellers $k$ are bounded, we can bypass the $\PPAD$-hardness of
Theorem~\ref{thm:bh-hardness} and design an efficient polynomial-time
approximation scheme (Theorem~\ref{thm:eptas-intro}). The key observation
behind our algorithm is that for any seller without captive buyers,
multiplying both its own price and all its neighbors' prices by a common
factor $\lambda \in [0,1]$ scales both its expected revenue and its
maximum deviation revenue by $\lambda$. This scale invariance allows us
to restrict computation to a bounded-radius neighborhood $B_R$ around the
$k$ captive sellers: we compute an exact equilibrium on a finite price
grid inside $B_R$, smooth each positive grid point into a uniform
distribution over the preceding grid interval so that nearby scaling
factors induce small total variation distance, and linearly scale prices
down to zero as the graph distance from the captive sellers increases to
$R$.

To formalize this approach, fix an integer grid size $K \geq 1$ and grid
step size $h = 1/K$. Given an exact mixed equilibrium $\bX$ of the finite
game restricted to the price grid $\cbr{0,h,2h,\ldots,1}$, let $\bF$
denote the continuous strategy profile obtained by replacing each
positive grid price $jh$ ($1 \leq j \leq K$) with an independent uniform
draw on the interval $[(j-1)h, jh]$ while keeping price zero at zero. Our
first lemma shows that this smoothing step eliminates positive-price
atoms while yielding a $2h$-equilibrium of the continuous game.

\begin{lemma}[Smoothing a finite price-grid equilibrium]
\label{lem:gridsmoothing}
Let $h=1/K$, and let $\bX$ be an exact mixed equilibrium of the finite
game restricted to the price grid $\cbr{0,h,2h,\ldots,1}$. Then the
smoothed profile $\bF$ is a $2h$-equilibrium of the continuous Bertrand
network.
\end{lemma}
\begin{proof}
Consider any two adjacent sellers and condition on their sampled grid
prices under $\bX$. If their grid prices are distinct, smoothing
preserves their strict price ordering with probability one. If they tie
at a positive grid price $jh$, their two independent uniform draws on
$[(j-1)h,jh]$ give each seller win probability $1/2$, which matches the
equal tie-breaking rule on the grid, whereas a tie at price zero yields
zero revenue under both $\bX$ and $\bF$. Thus, each seller's expected
share of every incident market is unchanged by smoothing. Because
smoothing reduces each seller's price by at most $h$, seller $i$'s
expected revenue under $\bF$ decreases by at most $hM_i$ relative to
$\bX$, so that $u_i(\bF) \geq u_i(\bX) - hM_i$.

Next, consider any continuous price deviation $p\in[0,1]$ by seller $i$
against the smoothed opponent profile $\bF_{-i}$, and let
$p^-=h\Flr{p/h}$ be the nearest grid point at or below $p$. Against any
opponent whose grid price under $\bX$ is strictly greater than $p^-$
(and hence at least $p^-+h$), the grid deviation $p^-$ wins the entire
shared market with certainty. Against any opponent whose grid price under
$\bX$ is at most $p^-$, the smoothed opponent price lies in $[0,p^-]$, so
deviating to $p \geq p^-$ wins zero positive-price demand against that
opponent. Since $p - p^- \leq h$, it follows that
\begin{align*}
 u_i(p,\bF_{-i})\leq u_i(p^-,\bX_{-i})+hM_i\leq u_i(\bX)+hM_i,
\end{align*}
where the second inequality uses the fact that $\bX$ is an exact
equilibrium on the grid. Combining this upper bound with
$u_i(\bF) \geq u_i(\bX) - hM_i$ shows that
$\sup_{p\in[0,1]} u_i(p,\bF_{-i}) - u_i(\bF) \leq 2hM_i$ for every seller
$i$.
\end{proof}

After smoothing the grid equilibrium, each seller's marginal distribution
$F_i$ is a mixture of a point mass at zero and uniform distributions on
the grid intervals $[(j-1)h,jh]$ for $1 \leq j \leq K$. For any scalar
factor $\lambda \in [0,1]$ and distribution $F$, let $\lambda F$ denote
the distribution of the rescaled price $\lambda P$ when $P \sim F$. In
the next lemma, we show that rescaling such a smoothed distribution $F$
by two nearby factors $\lambda$ and $\mu$ induces a small total variation
distance $d_{\mathrm{TV}}(\lambda F, \mu F)$.

\begin{lemma}[Total variation stability under price scaling]
\label{lem:scaleTV}
Let $F$ be a mixture of a point mass at zero and uniform distributions on
$[(j-1)h,jh]$ for $1\leq j\leq K$. Then for any $\lambda,\mu\in[0,1]$
with $\max\cbr{\lambda,\mu}>0$, we have
\begin{align}
 d_{\mathrm{TV}}(\lambda F,\mu F)
 \leq\min\cbr{1,\frac{\abs{\lambda-\mu}}{h\max\cbr{\lambda,\mu}}}.
 \label{eq:scaleTV}
\end{align}
\end{lemma}
\begin{proof}
Because a point mass at zero is invariant under scalar multiplication,
convexity of total variation distance implies that it suffices to prove
the bound when $F$ is the uniform distribution on a single grid interval
$[a,b]=[(j-1)h,jh]$ of length $b-a=h$. Assume without loss of generality
that $0<\lambda\leq\mu$. Under rescaling, $\lambda F$ and $\mu F$ are
uniformly distributed on $[\lambda a,\lambda b]$ and $[\mu a,\mu b]$,
respectively. When these two intervals overlap (that is, when
$\mu a < \lambda b$), the overlap region $[\mu a,\lambda b]$ carries
shared probability mass $(\lambda b-\mu a)/(\mu h)$, and hence
\begin{align*}
 d_{\mathrm{TV}}(\lambda F,\mu F)
 =1-\frac{\lambda b-\mu a}{\mu h}
 =\frac{(\mu-\lambda)b}{\mu h}
 \leq\frac{\mu-\lambda}{\mu h},
\end{align*}
where the inequality uses $b \leq 1$. When the two scaled intervals are
disjoint ($\mu a \geq \lambda b$), we have
$(\mu-\lambda)/(\mu h) \geq (\mu b-\lambda b)/(\mu b-\mu a) \geq 1$, so
the right-hand side of~\eqref{eq:scaleTV} equals $1 \geq d_{\mathrm{TV}}(\lambda F,\mu F)$.
\end{proof}

Because the smoothed and scaled distributions remain continuous, the
final step of our algorithm discretizes each seller's distribution into a
finite-support strategy by taking $K$ equal-probability quantiles. The
following lemma shows that approximating each seller's marginal CDF
within uniform distance $\theta$ increases the normalized equilibrium
error by at most $4\theta$.

\begin{lemma}[Finite-support quantile approximation]
\label{lem:quantile}
Suppose $\bF$ is a $\delta$-equilibrium and $\bG$ is a strategy profile
satisfying $\sup_{p\in[0,1]}\abs{F_i(p)-G_i(p)}\leq\theta$ for every
seller $i$. Then $\bG$ is a $(\delta+4\theta)$-equilibrium. 
In particular, discretizing each $F_i$ by placing probability $1/N$ at each generalized quantile $q_{i,j}=\inf\cbr{p\in[0,1]:F_i(p)\geq(j-1/2)/N}$ for $1\leq j\leq N$ (summing probabilities when quantile points coincide) achieves $\theta\leq 1/N$.
\end{lemma}
\begin{proof}
For any opponent with CDF $H$ and left-hand limit
$H(p^-)=\lim_{t\uparrow p} H(t)$, seller $i$'s probability of winning a
shared edge at price $p$ (including half-ties) is
$1-\rbr{H(p)+H(p^-)}/2$. Replacing each opponent's CDF $F_j$ by $G_j$
with $\sup_{p\in[0,1]}\abs{F_j(p)-G_j(p)}\leq\theta$ therefore changes
seller $i$'s expected revenue at any price $p\in[0,1]$ by at most
$\theta\sum_j\beta_{ij}\leq\theta M_i$. In particular, seller $i$'s
supremal deviation revenue changes by at most $\theta M_i$.

To bound the change in seller $i$'s own expected revenue when $F_i$ is
replaced by $G_i$, fix an opponent price $q\in[0,1]$. For a seller with
CDF $H$, integration by parts expresses the expected revenue from a
unit-mass shared edge against price $q$ as
\begin{align*}
 \Ex{P\Ind{P<q}+\tfrac12 P\Ind{P=q}}
 =\frac{q}{2}\rbr{H(q^-)+H(q)}-\int_0^q H(t)\,dt.
\end{align*}
Replacing $H=F_i$ by $G_i$ alters the boundary term by at most $q\theta$
and the integral by at most $q\theta$, changing the expected edge revenue
by at most $2q\theta\leq 2\theta$. Similarly, captive revenue
$\alpha_i\Ex{P_i}=\alpha_i\int_0^1(1-H(t))\,dt$ changes by at most
$\alpha_i\theta$. Integrating over the opponents' prices shows that
replacing $F_i$ by $G_i$ changes seller $i$'s expected revenue by at most
$2\theta M_i$, and replacing the opponents' distributions $\bF_{-i}$ by
$\bG_{-i}$ changes it by at most an additional $\theta M_i$. Combining
this $3\theta M_i$ bound on actual revenue with the $\theta M_i$ bound on
deviation revenue proves that $\bG$ is a $(\delta+4\theta)$-equilibrium.
\end{proof}

We summarize the complete approximation scheme in
Algorithm~\ref{alg:eptas} and now prove Theorem~\ref{thm:eptas-intro}.

\begin{pseudocode}{EPTAS via neighborhood smoothing and scaling}{alg:eptas}
\textbf{Input:} Network $(\balpha,\beta)$ and accuracy $\eps\in(0,1]$.
\begin{enumerate}
\item Let $\cC=\cbr{i:\alpha_i>0}$. If $\cC=\varnothing$, return price
zero at every seller.
\item Set $K=\Cel{8/\eps}$, $h=1/K$, and $R=\Cel{16/(\eps h)}$. Compute
graph distances $d_i=\operatorname{dist}(i,\cC)$ and form the induced
subnetwork on $B_R=\cbr{i:d_i\leq R}$.
\item Enumerate nonempty grid-support subsets of $\cbr{0,h,\ldots,1}$ for
sellers in $B_R$ and solve the linear feasibility system for an exact
grid equilibrium $\bX$ on $B_R$.
\item Smooth each positive grid price $jh$ into a uniform draw on
$[(j-1)h,jh]$ and scale seller $i$'s prices by $\lambda_i=1-d_i/R$ for
$i\in B_R$, setting price zero outside $B_R$.
\item Discretize each seller's scaled distribution at the $K$ quantile
levels $(j-1/2)/K$ ($1\leq j\leq K$) with equal probability $1/K$.
\end{enumerate}
\end{pseudocode}

\begin{proof}[Proof of Theorem~\ref{thm:eptas-intro}]
If $k=0$, assigning price zero to every seller is an exact equilibrium,
so assume $k>0$ and define the parameters
\begin{align}
 K=\Cel{8/\eps},\quad h=1/K,\quad
 R=\Cel{\frac{16}{\eps h}},\quad \kappa=1/R.
 \label{eq:eptasparameters}
\end{align}
Let $\bF$ denote the smoothed $2h$-equilibrium on the induced subnetwork
$B_R$ guaranteed by Lemma~\ref{lem:gridsmoothing}, and let $\bP$ denote
the rescaled strategy profile on the full graph under which each seller
$i\in B_R$ plays $\lambda_i F_i$ with $\lambda_i=1-d_i/R$ and each seller
outside $B_R$ plays price zero.

First, consider any seller $i$ with $d_i<R$. By definition of $B_R$, all
neighbors $j$ of $i$ in $G$ belong to $B_R$, so seller $i$ has the same
incident edges and total accessible mass $M_i$ in $B_R$ as in $G$.
Moreover, $\lambda_i\geq\kappa$ and $\abs{\lambda_i-\lambda_j}\leq\kappa$
for every neighbor $j$ of $i$. Compare seller $i$'s payoff against its
actual rescaled opponents $\lambda_j F_j$ with its payoff against
hypothetical opponents that all use the uniform scaling factor
$\lambda_i F_j$. By Lemma~\ref{lem:scaleTV}, for every neighbor $j$ we
have $d_{\mathrm{TV}}(\lambda_i F_j,\lambda_j F_j)\leq\min\cbr{1,\kappa/(\lambda_i h)}$.
Because no neighbor prices above $\lambda_i+\kappa$ under either profile,
any price $p>\lambda_i+\kappa$ earns zero shared-market revenue in both
cases, whereas for any price $p\leq\lambda_i+\kappa$, the expected
revenue of seller $i$ against the actual opponents $\lambda_j F_j$
differs from that against the uniformly scaled opponents $\lambda_i F_j$
by at most
\begin{align}
 (\lambda_i+\kappa)\min\cbr{1,\frac{\kappa}{\lambda_i h}}\sum_j\beta_{ij}
 \leq\frac{2\kappa}{h}M_i.
 \label{eq:scalingpayoff}
\end{align}
Notice that if $d_i>0$, seller $i$ has no captive buyers ($\alpha_i=0$),
so rescaling both seller $i$ and all its neighbors by the common factor
$\lambda_i$ scales both its expected revenue and its supremal deviation
revenue by $\lambda_i$, which preserves the $2hM_i$ deviation bound from
Lemma~\ref{lem:gridsmoothing}. If $d_i=0$ (so seller $i$ is a captive
seller), then $\lambda_i=1$ and no rescaling occurs at $i$. In either
case, replacing the uniformly scaled opponents $\lambda_i F_j$ by the
actual opponents $\lambda_j F_j$ changes both seller $i$'s actual
expected revenue and its supremal deviation revenue by at
most~\eqref{eq:scalingpayoff}, so seller $i$'s maximum deviation gain
under $\bP$ is at most
\begin{align*}
 \rbr{2h+\frac{4\kappa}{h}}M_i\leq\frac{\eps}{2}M_i.
\end{align*}

Second, consider any seller $i$ with $d_i\geq R$. Such a seller plays
price zero, has no captive buyers, and has only neighbors with scaling
factors $\lambda_j\leq\kappa$. Consequently, deviating to any price
$p>\kappa$ wins zero buyers, while deviating to $p\leq\kappa$ earns at
most $\kappa M_i\leq(\eps/2)M_i$. Thus, $\bP$ is an
$(\eps/2)$-equilibrium on the entire graph $G$. Applying the $K$-quantile
discretization of Step~5 increases the normalized equilibrium error by at
most $4/K=4h\leq\eps/2$ by Lemma~\ref{lem:quantile}, yielding a
finite-support $\eps$-equilibrium.

Finally, to bound the running time, observe that the radius-$R$
neighborhood around the $k$ captive sellers contains at most
$\abs{B_R}\leq k(\Delta+1)^{R+1}$ vertices, where $R=O(\eps^{-2})$. In
Step~3, there are at most $2^{(K+1)\abs{B_R}}$ choices for the joint grid
supports of the sellers in $B_R$. For each fixed support profile, finding
a grid distribution $\bX$ satisfying the support, indifference, and
best-response constraints reduces to solving a linear feasibility program
of size $\poly(\abs{B_R},K,\cL)$. Since at least one support profile is
feasible by Nash's theorem, Step~3 finds an exact grid equilibrium in
time
\begin{align*}
 2^{(K+1)\abs{B_R}}\poly(\cL)
 =2^{O\rbr{k\eps^{-1}(\Delta+1)^{O(\eps^{-2})}}}\poly(\cL),
\end{align*}
and Steps~4 and~5 compute the rational quantile prices in polynomial time.
\end{proof}

\section{Structural Complexity}
\label{sec:finite}

We now prove Theorem~\ref{thm:finite} by fixing an achievable equilibrium
revenue vector $\bu$ and analyzing the space of marginal price
distributions that realize $\bu$. Let $B$ denote the symmetric matrix
with entries $B_{ij}=\beta_{ij}$ on edges and $B_{ii}=0$. Recall from~\cite[Theorem~4.5]{BLN13}
that connected components of the graph without captive buyers earn zero
revenue in every equilibrium and can be assigned price zero. On the
remaining components, let $n$ be the number of sellers. Every such seller
$i$ earns strictly positive revenue $u_i>0$ because captive sellers can
guarantee $\alpha_i>0$ by posting price one, and any neighbor $j$ can
undercut seller $i$'s minimum support price $u_i/M_i$.

\subsection{Convex Reformulation}

Choose a constant $H\geq 2\max_i M_i/u_i$, set $T=H-1$ and $\lambda=1/H$,
and apply the reciprocal-price transformation
\begin{align}
 X_i=P_i^{-1}-1,\quad
 U_i(t)=\Pr{X_i\leq t}=\Pr{P_i\geq(1+t)^{-1}}.
 \label{eq:finite-transform}
\end{align}
Because no seller posts a price below $u_i/M_i\geq 2\lambda$ in
equilibrium, we have $U_i(t)=1$ for all $t$ near $T$, while
$U_i(0)=\Pr{P_i=1}$ records any point mass at price one.

For any price $p=(1+t)^{-1}>0$, undercutting $p$ by an arbitrarily small
amount captures the entire shared market against every neighbor charging
at least $p$, which yields the necessary equilibrium condition
\begin{align}
 p\rbr{\alpha_i+\sum_j B_{ij}\Pr{P_j\geq p}}\leq u_i.
 \label{eq:finite-undercut}
\end{align}
Motivated by~\eqref{eq:finite-undercut}, we say that a profile of
right-continuous functions $\bU=(U_1,\ldots,U_n)$ on $[0,T]$ is
\emph{feasible} if
\begin{align}
 \begin{gathered}
 U_i\text{ is nondecreasing},\quad
 \mathbf{0}\leq\bU(t)\leq\bone,\quad\bU(T)=\bone,\\
 B\bU(t)\leq\bu(1+t)-\balpha\quad\text{for all } t\in[0,T].
 \end{gathered}
 \label{eq:finite-feasible}
\end{align}
By construction, every equilibrium realizing $\bu$ induces a feasible
profile $\bU$. Conversely, any feasible profile $\bU$ defines independent
price distributions supported on $[\lambda,1]$ under which no seller can
earn more than $u_i$ by deviating to any price $p\in[0,1]$, since any
price $p<\lambda$ earns at most $\lambda M_i\leq u_i/2$.

To ensure that each seller's expected revenue under $\bU$ actually equals
$u_i$, let $d_i(p)$ denote the expected buyer mass won by seller $i$ at
price $p$ (including half of any ties), and let
$\cA=\sum_i\alpha_i+\sum_{\cbr{i,j}\in E}\beta_{ij}$ denote the total
buyer mass in the network. Since every buyer purchases from some seller,
we have $\sum_i\Ex{d_i(P_i)}=\cA$. Moreover, integration by parts
expresses the revenue-weighted sum of expected reciprocal prices as
\begin{align}
 \Phi(\bU)\defeq\sum_i u_i\Ex{P_i^{-1}}
 =H\sum_i u_i-\int_0^T\bu^\top\bU(t)\,dt.
 \label{eq:finite-objective}
\end{align}
Subtracting $\cA=\sum_i\Ex{d_i(P_i)}$ from $\Phi(\bU)$ shows that every
feasible profile $\bU$ satisfies the deficit identity
\begin{align}
 \Phi(\bU)-\cA
 =\sum_i\Ex{\frac{u_i-P_i d_i(P_i)}{P_i}}\geq 0,
 \label{eq:finite-gap}
\end{align}
where nonnegativity holds because $p d_i(p)\leq u_i$ for all
$p\in[\lambda,1]$. Consequently, equality holds in~\eqref{eq:finite-gap}
if and only if $P_i d_i(P_i)=u_i$ almost surely for every seller $i$,
which is equivalent to $\bU$ forming an exact equilibrium with revenue
vector $\bu$. Since $\bu$ is achievable, the exact equilibria realizing
$\bu$ are precisely the feasible profiles $\bU$ that maximize the linear
integral objective $\int_0^T\bu^\top\bU(t)\,dt$ over~\eqref{eq:finite-feasible}.

Finally, if we define the nonnegative slack functions
\begin{align}
 r_i(t)=u_i(1+t)-\alpha_i-(B\bU(t))_i\geq 0,
 \label{eq:finite-slack}
\end{align}
then $(u_i-P_i d_i(P_i))/P_i\geq r_i(X_i)$ almost surely. Thus, every
equilibrium realizing $\bu$ satisfies the complementary slackness
condition
\begin{align}
 \int_{[0,T]}r_i(t)\,dU_i(t)=0\quad\text{for all } i\in[n].
 \label{eq:finite-complementarity}
\end{align}

\subsection{Bounding via Concavity}

For an equilibrium $\bU$ realizing $\bu$, consider the scalar function
$V(t)=\bu^\top\bU(t)$. On any subinterval $[a,b]\subseteq[0,T]$,
replacing $\bU(t)$ by the linear interpolation between its endpoint
values $\bU(a)$ and $\bU(b)$ preserves all feasibility conditions
in~\eqref{eq:finite-feasible}. Because $\bU$ maximizes
$\int_0^T V(t)\,dt$, we must have
\begin{align}
 \int_a^b V(t)\,dt\geq\frac{b-a}{2}\rbr{V(a)+V(b)}.
 \label{eq:finite-chord}
\end{align}
Dividing both sides of~\eqref{eq:finite-chord} by $b-a$ and taking the
limit $a\uparrow b$ rules out upward jumps at any $b>0$. Combined with
right-continuity, this implies that $V$ (and hence each $U_i$, since
$\bu>\mathbf{0}$) is continuous on $[0,T]$. Furthermore,~\eqref{eq:finite-chord}
implies that $V$ is concave on $[0,T]$, which ensures that $V$, each
$U_i$, and each slack function $r_i$ are locally Lipschitz and
absolutely continuous on $(0,T)$.

At almost every $t\in(0,T)$, both $\bU$ and $\br$ are differentiable and
satisfy~\eqref{eq:finite-complementarity}. Let $v_i=U_i'(t)\geq 0$ and
define the active seller set $S=\cbr{i\in[n]:v_i>0}$. For every $i\in S$,
complementary slackness forces $r_i(t)=0$, and since $r_i\geq 0$ attains
a local minimum at $t$, we also have $r_i'(t)=0$. Differentiating
$r_i(t)=0$ for $i\in S$ yields the linear system
\begin{align}
 B_{SS}\bv_S=\bu_S,\quad \bv_{\overline{S}}=\mathbf{0}.
 \label{eq:finite-rates}
\end{align}
Even when the principal submatrix $B_{SS}$ is singular, any two solutions
$\bw_S$ and $\bz_S$ to $B_{SS}\bv_S=\bu_S$ satisfy
\begin{align*}
 \bu_S^\top(\bw_S-\bz_S)=\bz_S^\top B_{SS}(\bw_S-\bz_S)=0
\end{align*}
by the symmetry of $B_{SS}$. Thus, each active subset $S\subseteq[n]$
induces at most one possible scalar slope $V'(t)=\bu_S^\top\bv_S$, so
$V'(t)$ takes values almost everywhere in a finite set $\Lambda(B,\bu)$
of size at most $2^n$. Because $V$ is concave, its nonincreasing
derivative $V'$ can visit each value in $\Lambda(B,\bu)$ on at most one
interval, which proves that $V$ is piecewise affine on $[0,T]$ with at
most $2^n$ intervals.

\subsection{Linear Interpolation}

We now complete the proof of Theorem~\ref{thm:finite} and derive two
structural corollaries.

\begin{proof}[Proof of Theorem~\ref{thm:finite}]
Let $0=t_0<t_1<\cdots<t_K=T$ ($K\leq 2^n$) be the partition of $[0,T]$
into maximal intervals on each of which $V(t)=\bu^\top\bU(t)$ is affine.
On each interval $[t_{k-1},t_k]$, replace the vector function $\bU(t)$ by
the linear interpolation between $\bU(t_{k-1})$ and $\bU(t_k)$. By
convexity of the constraints in~\eqref{eq:finite-feasible}, the
interpolated profile remains feasible. Moreover, because $V(t)=\bu^\top\bU(t)$
is already affine on $[t_{k-1},t_k]$, linear interpolation leaves $V(t)$
unchanged at every point $t\in[0,T]$, preserving the optimal objective
value $\int_0^T V(t)\,dt$. By the deficit identity~\eqref{eq:finite-gap},
the interpolated profile is an exact equilibrium with the same revenue
vector $\bu$. Transforming back to price coordinates via
$F_i(p)=1-U_i(p^{-1}-1)$ for $p\in(0,1)$ and assigning point mass
$U_i(0)$ at $p=1$ yields a reciprocal-affine equilibrium on at most
$2^n$ intervals.
\end{proof}

The same variational characterization immediately implies two structural
corollaries.

\begin{corollary}[Convexity at fixed revenues]
\label{cor:fiber-convex}
For any achievable positive revenue vector $\bu$, the set of equilibrium
marginal CDF profiles realizing $\bu$ is convex.
\end{corollary}
\begin{proof}
Any convex combination of two feasible profiles $\bU^{(1)}$ and
$\bU^{(2)}$ maximizing $\int_0^T\bu^\top\bU(t)\,dt$ remains feasible
in~\eqref{eq:finite-feasible} and attains the same maximum objective, so
by~\eqref{eq:finite-gap} it forms an equilibrium with revenue vector
$\bu$.
\end{proof}

\begin{corollary}[Uniqueness of the revenue-weighted price distribution]
\label{cor:scalar-unique}
For any achievable positive revenue vector $\bu$, the aggregate scalar
function $V(t)=\sum_i u_i U_i(t)$, and equivalently the weighted measure
$\sum_i u_i\operatorname{Law}(P_i)$, is identical across all equilibria
realizing $\bu$.
\end{corollary}
\begin{proof}
Let $V_1$ and $V_2$ correspond to two equilibria realizing $\bu$, and pick
a mixing weight $\theta\in(0,1)$ outside the finite set
$\cbr{(c-b)/(a-b):a,b,c\in\Lambda(B,\bu),\ a\neq b}$. By
Corollary~\ref{cor:fiber-convex}, $\theta V_1+(1-\theta)V_2$ also has its
derivative in $\Lambda(B,\bu)$ almost everywhere, which forces
$V_1'=V_2'$ almost everywhere and hence $V_1=V_2$ on $[0,T]$.
\end{proof}

\section{Tightness of Structural Complexity}
\label{sec:exp-lower}

We now prove Theorem~\ref{thm:exp-lower} by constructing a family of
 Bertrand networks on complete graphs and prescribed 
revenue vectors $\bu$ whose unique realizing equilibrium has
$3\cdot 2^k-2$ reciprocal-affine pieces.

\subsection{Steepest Ascent on Concave Polyhedra}

Let $f(\by)=\min_i(\bp_i^\top\by+b_i)$ be the pointwise minimum of
finitely many affine functions. 
At each point $\by$, let $\bg(\by)$
denote the unique minimum-norm vector in the convex hull of the gradients
$\bp_i$ of the planes active at $\by$, and consider the continuous steepest-ascent
trajectory $\by:[0,\infty)\to\R^d$ satisfying
\begin{align*}
 \dot{\by}(t)=\bg(\by(t)),
\end{align*}
where time derivatives are taken from the right.

The following lemma
summarizes the piecewise-affine structure and $\ell_2$-stability of this
flow.

\begin{lemma}[Polyhedral ascent and stability]
\label{lem:lb-flow}
From any initial point, the trajectory $\by(t)$ exists uniquely, is
piecewise affine with finitely many velocity changes, and has strictly
decreasing squared speed $\norm{\bg(\by(t))}^2$ across velocity changes.
Its value function $f(\by(t))$ is continuous, concave, and piecewise
affine with derivative $\norm{\bg(\by(t))}^2$. Moreover, if $f$ and
$\widehat{f}$ differ uniformly by at most $\epsilon$ on a domain
containing their trajectories $\by(t)$ and $\widehat{\by}(t)$ over
$t\in[0,1]$ from a common starting point, then
\begin{align}
 \norm{\by(t)-\widehat{\by}(t)}\leq 2\sqrt{\epsilon t}
 \quad\text{for all } t\in[0,1],
 \label{eq:lb-flow-stability}
\end{align}
and if $f$ is $1$-Lipschitz, $\abs{f(\by(t))-\widehat{f}(\widehat{\by}(t))}\leq 2\sqrt{\epsilon}+\epsilon$.
\end{lemma}
\begin{proof}
By standard convex projection properties, $\bp_i^\top\bg(\by)\geq\norm{\bg(\by)}^2$
for all active planes $i$, with equality for every plane receiving
positive convex weight in $\bg(\by)$. Moving in direction $\bg(\by)$
increases $f$ at rate $\norm{\bg(\by)}^2$ and maintains the active set
until a previously inactive plane $j$ first ties for the minimum. At that
instant, $\bp_j^\top\bg(\by)<\norm{\bg(\by)}^2$, so including $\bp_j$ in
the active convex hull strictly reduces the minimum norm. Because there
are finitely many subsets of planes, the trajectory has finitely many
affine pieces.

For stability, the supergradient inequality for concave functions gives
almost everywhere
\begin{align*}
 \frac12\frac{d}{dt}\norm{\by-\widehat{\by}}^2
 =(\by-\widehat{\by})^\top(\bg(\by)-\widehat{\bg}(\widehat{\by}))
 \leq f(\by)-f(\widehat{\by})+\widehat{f}(\widehat{\by})-\widehat{f}(\by)
 \leq 2\epsilon.
\end{align*}
Integrating from $0$ to $t$ yields~\eqref{eq:lb-flow-stability}, which in
turn implies uniqueness when $\epsilon=0$.
\end{proof}

\subsection{Inductive Doubling of Speed Phases}

We inductively construct a concave polyhedral function $f_k$ on
$\R^{m_k}$ with $m_k=2k+1$ planes whose gradients have norm at most one
and intercepts lie in $[0,1]$, starting at $\by(0)=\mathbf{0}$ with
$f_k(\mathbf{0})=0$ and constant velocity after $t=1$. For the base case
$k=0$, we set $f_0(y)=y$.

Given $f=f_{k-1}$ on $\R^{m_{k-1}}$, define its terminal quantities at
time one by
\begin{align*}
 \bd=\by(1),\quad F=f(\bd),\quad \bg_*=\lim_{t\downarrow 1}\frac{d\by(t)}{dt},\quad
 r_0=\norm{\bd}^2,\quad q_*=\norm{\bg_*}^2,
\end{align*}
where $\norm{\bd},F,q_*\in[0,1]$. Introducing two new scalar coordinates
$(z,w)$, we define the unnormalized function $\cF$ and its rescaled
counterpart $f_k$ by
\begin{align}
 \begin{aligned}
 \cF(\by,z,w)&=\min\cbr{f(\by)+10z,\ -\bd^\top\by+5z+2w+F+r_0+50,\ w+F+r_0+127},\\
 f_k(\bY)&=\frac{\cF(1500\bY)}{18000}.
 \end{aligned}
 \label{eq:lb-doubling}
\end{align}
Denote the two newly added planes in $\cF$ by $B$ and $C$. In the next
lemma, we show that the steepest-ascent trajectory of $\cF$ traverses the
trajectory of $f$, resets the $\by$-coordinates to zero along plane $B$,
waits on plane $C$, and then replays the entire trajectory of $f$ at a
slower speed.

\begin{lemma}[Four-stage trajectory of $\cF$]
\label{lem:lb-doubling}
Starting from $(\mathbf{0},0,0)$, the steepest-ascent trajectory of $\cF$
proceeds through four consecutive stages:
\begin{enumerate}
\item For $t\in[0,1]$, it follows $(\by(t),10t,0)$ to $(\bd,10,0)$.
\item For $t\in[1,2]$, it follows velocity $(-\bd,5,2)$ along plane $B$
to $(\mathbf{0},15,2)$.
\item For the next $21-r_0-F$ time units, it follows velocity
$(\mathbf{0},0,1)$ along plane $C$ to $(\mathbf{0},15,23-r_0-F)$.
\item It replays the trajectory $(\by(s),15+10s,23-r_0-F+100s+f(\by(s)))$
for $s\in[0,1]$ under the time reparameterization
\begin{align}
 \frac{dt}{ds}=101+\norm{\bg(\by(s))}^2,
 \label{eq:lb-replay-time}
\end{align}
reaching $s=1$ at total time $124-r_0$ and continuing at constant
velocity $\frac{1}{101+q_*}(\bg_*,10,q_*+100)$ thereafter.
\end{enumerate}
\end{lemma}
\begin{proof}
During Stage~1, the gap between plane $B$ and the active old minimum
$f(\by)+10z$ has time derivative $-\bd^\top\bg-\norm{\bg}^2-50<0$ and
reaches zero precisely at $t=1$, whereas plane $C$ remains at least $27$
above the minimum. At $t=1$, every active old plane has gradient
$\ba=(\bp_i,10,0)$ satisfying
\begin{align*}
 \ba^\top(-\bd,5,2)\geq -1+50=49 > r_0+29=\norm{(-\bd,5,2)}^2,
\end{align*}
so plane $B$ alone determines the velocity $(-\bd,5,2)$ during Stage~2,
while all old planes separate from $B$ at rate at least $19$.

After one unit of time in Stage~2, the trajectory reaches $(\mathbf{0},15,2)$,
where plane $C$ meets plane $B$ at value $F+r_0+129$. Since
$(\mathbf{0},0,1)^\top(-\bd,5,2)=2>1$, the velocity switches to the
gradient $(\mathbf{0},0,1)$ of plane $C$ alone. Because the old minimum
equals $f(\mathbf{0})+150=150$ at $(\mathbf{0},15,2)$, Stage~3 lasts for
$150-(F+r_0+129)=21-r_0-F\geq 19$ time units while $B-C$ increases.

When $C$ meets the old minimum at $(\mathbf{0},15,23-r_0-F)$, minimizing
the squared norm of a convex combination $\theta(\bg,10,0)+(1-\theta)(\mathbf{0},0,1)$
yields
\begin{align}
 \theta=\frac{1}{101+\norm{\bg}^2},
 \quad\text{so}\quad
 \rbr{\frac{d\by}{dt},\frac{dz}{dt},\frac{dw}{dt}}=(\theta\bg,10\theta,1-\theta).
 \label{eq:lb-replay-velocity}
\end{align}
Thus $ds/dt=\theta$, which replays the old trajectory $\by(s)$ according
to~\eqref{eq:lb-replay-time} while plane $B$'s gap increases at rate
$\theta(-\bd^\top\bg+\norm{\bg}^2+150)>0$. Integrating $dt/ds$ over
$s\in[0,1]$ gives duration $101+\int_0^1\norm{\bg}^2\,ds=101+F$, reaching
$(\bd,25,123-r_0)$ at total time $124-r_0$, after which the velocity
remains constant.
\end{proof}

Rescaling by~\eqref{eq:lb-doubling} divides all gradients by $12$ and
compresses $125$ time units into $[0,1]$. Thus, the number of distinct
squared-speed intervals $K_k$ on $[0,1]$ satisfies
\begin{align}
 K_0=1,\quad K_k=2K_{k-1}+2=3\cdot 2^k-2.
 \label{eq:lb-piece-count}
\end{align}
Moreover, an inductive check of the map $a\mapsto(a+100)/144$ in Stage~1
and $a\mapsto(a+100)/(144(a+101))$ in Stage~4 shows that every speed
interval of $f_k$ has length at least $\ell_k$ and consecutive squared
speeds differ by at least $\Delta_k$, where
\begin{align}
 \ell_k=125^{-k},\quad
 \Delta_k=(2\,000\,000)^{-k}.
 \label{eq:lb-separation}
\end{align}

\subsection{Polynomial-time Rounding}

To keep the bit complexity of the construction polynomial in $k$, let
$P_k$ denote the $m\times m$ matrix of plane gradients of $f_k$, where
$m=2k+1$. Because $\bd=P_{k-1}^\top\ba$ for a probability vector $\ba$,
we have $\bd^\top P_{k-1}^{-1}\bone=1$, so the unnormalized block matrix
in~\eqref{eq:lb-doubling} has determinant $15\det P_{k-1}$. After
dividing by $12$, we obtain
\begin{align}
 \det P_k=\frac{15^k}{12^{k^2+2k}},\quad
 \sigma_{\min}(P_k)\geq s_k\defeq 12^{-(k^2+2k)}m^{-m}.
 \label{eq:lb-conditioning}
\end{align}

We choose the largest dyadic mesh $\delta>0$ satisfying
\begin{align}
 \delta\leq\min\cbr{\frac{s_k}{4m},\ \frac{(\Delta_k\ell_k)^2}{10^8 m},\ \frac{1}{10m}}
 \label{eq:lb-rounding-conditions}
\end{align}
and round every coefficient of $f_k$ to an integer multiple of $\delta$,
obtaining a polyhedral function $\widehat{f}$ with gradient
matrix $P$ satisfying $\sigma_{\min}(P)\geq s_k/2$. By
Lemma~\ref{lem:lb-flow}, the value histories of $f_k$ and $\widehat{f}$
differ on $[0,1]$ by at most $3\sqrt{3m\delta}<\Delta_k\ell_k/64$.
Comparing secant slopes of length $\ell_k/4$ around the midpoint of each
ideal speed piece shows that $\widehat{f}$ retains at least $K_k$
distinct squared-speed pieces on $[0,1]$.

To compute the rounded coefficients of $f_k$ without simulating all $K_k$
pieces, set
\begin{align*}
 \bG=\frac{(\bg_*,10,q_*+100)}{q_*+101},\quad
 Q=\frac{q_*+100}{q_*+101}.
\end{align*}
Lemma~\ref{lem:lb-doubling} implies that the terminal state at $t=1$ obeys
the exact algebraic recurrence
\begin{align}
 \bd_k=\frac{(\bd,25,123-r_0)+(1+r_0)\bG}{1500},\quad
 F_k=\frac{250+F+(1+r_0)Q}{18000},\quad
 \bg_{*,k}=\frac{\bG}{12}.
 \label{eq:lb-endpoint-recurrence}
\end{align}
Evaluating~\eqref{eq:lb-endpoint-recurrence} over $k$ steps using
interval arithmetic with $O(k^2+\log(1/\delta))=\poly(k)$ bits of
precision computes the rounded coefficients of $\widehat{f}$ in
$\poly(k)$ time.

\subsection{Conversion to a Twin Bertrand Network}
\begin{proof}[Proof of Theorem~\ref{thm:exp-lower}]
For each $i\in[r]$, we create twin sellers $i^+,i^-$, each with captive
mass $\alpha_i'$ and target revenue $u_i'$, connected by a mutual edge of
mass $\widetilde{C}_{ii}$ and cross-edges of mass $\widetilde{C}_{ij}/2$
for $i\neq j$. Assigning both twins the atomless survival function
$\Pr{P_{i^\pm}\geq(1+t)^{-1}}=Y_i(t)$ defines an exact equilibrium on the
complete graph of $N=2r=4k+4$ sellers with revenue vector $\bu$. By
construction, its revenue-weighted scalar invariant $V(t)$ satisfies
\begin{align}
 V(t(\tau))=2c\tau,\quad
 \frac{dt}{d\tau}=\frac{\rho-q(\tau)}{c},\quad
 V'(t)=\frac{2c^2}{\rho-q(\tau)}.
 \label{eq:lb-weighted-slopes}
\end{align}
Because $q(\tau)$ takes at least $K_k=3\cdot 2^k-2$ distinct values on
$\tau\in[0,1]$, the function $V(t)$ has at least $3\cdot 2^k-2$ affine
pieces and $3\cdot 2^k-3$ breakpoints in $(0,T)$, each of which must lie
in the support boundary of at least one seller.

It remains to show that this equilibrium profile is the unique
equilibrium realizing $\bu$. For any vector $\bx\in\R^{2r}$ on the twin
sellers, write $a_i=\zeta_i x_{i^+}$, $b_i=\zeta_i x_{i^-}$,
$m_i=(a_i+b_i)/2$, and $d_i=(a_i-b_i)/2$. Then the twin adjacency matrix
$B$ satisfies
\begin{align}
 \bx^\top B\bx
 =2\rho(\bone^\top\bm)^2-2\bm^\top D\bm-2\sum_{i=1}^r C_{ii}d_i^2,
 \quad\text{while}\quad
 \bu^\top\bx=2c\,\bone^\top\bm.
 \label{eq:lb-negative-subspace}
\end{align}
Since $D\succ 0$ and $C_{ii}>0$, the quadratic form $\bx^\top B\bx$ is
strictly negative definite on the subspace $\bu^\perp$. Now suppose that
$\bU$ and $\widehat{\bU}$ are two equilibria realizing $\bu$ with
respective slack vectors $\br$ and $\widehat{\br}$.
Corollary~\ref{cor:scalar-unique} implies that
$\bx(t)\defeq\bU(t)-\widehat{\bU}(t)\in\bu^\perp$ for all $t\in[0,T]$ and
$\bx(0)=\mathbf{0}$. By complementary slackness, we have almost
everywhere
\begin{align*}
 \frac{d}{dt}\rbr{\bx(t)^\top B\bx(t)}
 =-2(\br-\widehat{\br})^\top(\bU'-\widehat{\bU}')
 =2\br^\top\widehat{\bU}'+2\widehat{\br}^\top\bU'\geq 0.
\end{align*}
Since $\bx(0)^\top B\bx(0)=0$ and $\bx(t)^\top B\bx(t)\leq 0$ on
$\bu^\perp$, it follows that $\bx(t)=\mathbf{0}$ for all $t\in[0,T]$,
establishing uniqueness and completing the proof of
Theorem~\ref{thm:exp-lower}.
\end{proof}

\section{Doubly Exponential Algorithm}
\label{sec:exact-algorithm}

Combining the $2^n$ structural complexity bound of Theorem~\ref{thm:finite} with standard algorithms for solving polynomial systems over the reals~\citep{BPR96,BPR06} yields an exact algorithm for general Bertrand
networks.

\begin{theorem}[Exact computation]
\label{thm:exact-algorithm}
For any Bertrand network with $n$ sellers and input bit length
$\cL$, an exact equilibrium can be computed in time
\begin{align*}
 2^{2^{O(n)}}\poly(\cL).
\end{align*}
The algorithm outputs real-algebraic reciprocal-price breakpoints,
endpoint CDF values, and equilibrium revenues, and can also decide
whether a prescribed  revenue vector $\bu$ is achievable.
\end{theorem}
We provide the proof of this result in what follows.

\subsection{Polynomial Formulation over a Horizon}

After assigning price zero to connected components of the graph without
captive buyers, choose a captive root $r$ and a spanning tree in each
remaining component, and define positive  revenue lower bounds
inductively along each tree by
\begin{align}
 \ell_r=\alpha_r,\quad
 \ell_j=\frac{\beta_{ij}\ell_i}{2M_i}
 \quad\text{when $i$ is the parent of $j$}.
 \label{eq:exact-floor}
\end{align}
Because the root can charge price one and each child $j$ can undercut its
parent's support floor $u_i/M_i$ by charging $u_i/(2M_i)$, every
equilibrium satisfies $u_i\geq\ell_i$. Thus, setting
\begin{align}
 \lambda=\frac12\min_i\frac{\ell_i}{M_i},\quad
 H=1/\lambda,\quad T=H-1
 \label{eq:exact-horizon}
\end{align}
provides a common reciprocal-price horizon $T$ of bit length
$\poly(\cL)$ valid for all equilibria.

Set $N=2^n$ and introduce variables $u_i$, breakpoints
$0=t_0<t_1<\cdots<t_N=T$, and endpoint CDF values $z_{ik}=U_i(t_k)$.
Writing $r_{ik}=u_i(1+t_k)-\alpha_i-\sum_j B_{ij}z_{jk}$ and
$\Delta z_{ik}=z_{ik}-z_{i,k-1}$, we form the polynomial system
\begin{align}
 \begin{aligned}
 &\max\cbr{\alpha_i,\ell_i}\leq u_i\leq M_i,
      &&i\in[n],\\
 &0\leq z_{i0}\leq z_{i1}\leq\cdots\leq z_{iN}=1,
      &&i\in[n],\\
 &r_{ik}\geq 0,
      &&i\in[n],\ 0\leq k\leq N,\\
 &\Delta z_{ik}r_{i,k-1}=\Delta z_{ik}r_{ik}=0,
      &&i\in[n],\ 1\leq k\leq N,\\
 &z_{i0}(u_i-\alpha_i)=0,
      &&i\in[n],\\
 &z_{i0}z_{j0}=0,
      &&\cbr{i,j}\in E.
 \end{aligned}
 \label{eq:exact-system}
\end{align}
In the next lemma, we verify that solutions to~\eqref{eq:exact-system}
characterize exact reciprocal-affine equilibria on $N$ intervals.

\begin{lemma}[Correctness of the polynomial system]
\label{lem:exact-system}
A tuple of revenues, strictly increasing breakpoints, and endpoint values
satisfies~\eqref{eq:exact-system} if and only if linear interpolation of
$z_{ik}$ defines an exact reciprocal-affine equilibrium. Moreover, every
achievable revenue vector $\bu$ is realized by at least one solution.
\end{lemma}
\begin{proof}
Linearly interpolating $z_{ik}$ on each interval $[t_{k-1},t_k]$ yields
nondecreasing continuous functions $U_i$ on $[0,T]$ that satisfy
$B\bU(t)\leq\bu(1+t)-\balpha$ everywhere by convexity of the linear
constraints. Whenever $\Delta z_{ik}>0$, the complementarity equations
force $r_{i,k-1}=r_{ik}=0$ and hence $r_i(t)=0$ throughout $[t_{k-1},t_k]$,
so all interior prices in the support earn $u_i$. If $z_{i0}>0$, the last
two equations ensure $u_i=\alpha_i$ and forbid price-one atoms at
neighbors, so price one also earns $u_i$. The converse follows directly
by subdividing the $2^n$-piece representative of Theorem~\ref{thm:finite}
to $N=2^n$ intervals.
\end{proof}
\begin{pseudocode}{Exact equilibrium computation}{alg:exact}
\textbf{Input:} Market masses $(\balpha,\beta)$, and optionally a target revenue vector $\bu$.
\begin{enumerate}
\item If a prescribed $\bu$ has a positive entry in a connected component
      without captive buyers, report infeasibility. Otherwise assign
      price zero to all such components.
\item Compute the lower bounds $\ell_i$ and horizon $T$
      via~\eqref{eq:exact-floor}--\eqref{eq:exact-horizon}.
\item Set $N=2^n$ and form the system~\eqref{eq:exact-system} together
      with $0=t_0<t_1<\cdots<t_N=T$ (fixing $\bu$ if prescribed).
\item Run algebraic sampling~\citep{BPR96} on the resulting
      polynomial system. If no solution exists, report infeasibility.
\item Output the breakpoints $t_k$, endpoint values $z_{ik}$,
      and interpolated marginal CDFs $F_i(p)=1-U_i(p^{-1}-1)$ with atom
      $z_{i0}$ at price one.
\end{enumerate}
\end{pseudocode}

\subsection{Complexity Analysis}

\begin{proof}[Proof of Theorem~\ref{thm:exact-algorithm}]
By Lemma~\ref{lem:exact-system}, Algorithm~\ref{alg:exact} correctly
characterizes an exact reciprocal-affine equilibrium (or detects
infeasibility when a target revenue vector $\bu$ is prescribed). The
polynomial system~\eqref{eq:exact-system} contains $q=O(n2^n)$ variables
and $s=O(n2^n)$ polynomial equations and inequalities of degree $d\leq 3$,
each with rational coefficients of bit length $\poly(\cL)$. Applying the
real-algebraic sampling algorithm of~\cite{BPR96,BPR06} to a system of $s$
polynomials of degree at most $d$ in $q$ variables requires
$s^{q+1}d^{O(q)}=2^{2^{O(n)}}$ arithmetic operations on numbers of bit
length $2^{O(n)}\poly(\cL)$, and outputs an exact solution whose
coordinates are represented as roots of integer polynomials with
isolating rational intervals. Consequently, the total running time is
$2^{2^{O(n)}}\poly(\cL)$.
\end{proof}

\section{Open Questions}
\label{sec:questions}

Our results leave three natural directions for future work. First, while
Theorem~\ref{thm:exp-lower} shows that realizing a \emph{prescribed}
revenue vector can require exponentially many support intervals, does
every Bertrand network admit \emph{some} equilibrium whose
reciprocal-affine representation has polynomial size when revenues are
unconstrained? Second, can the doubly exponential runtime of
Theorem~\ref{thm:exact-algorithm} be improved?
Third, on generic sparse graphs, does \emph{every} equilibrium have
finitely many support boundaries, as asked by~\cite{BLN13}?

\section*{AI Use Statement}
The authors obtained initial partial results in early 2026. Subsequently, the full proofs were developed through extensive interaction between the authors and OpenAI Astra, with the authors providing key suggestions and critically verifying Astra's responses across iterative exchanges.
OpenAI Astra was also used in drafting and revising the manuscript via detailed prompts from the authors, with Gemini Argon assisting in editing parts of the text.
Lean formalization can be found in~\href{https://github.com/hajiaghayi/Bertrand
}{link}.

\bibliographystyle{plainnat}
\bibliography{ref}
\clearpage

\appendix

\section{Parameters for the hardness reduction}
\label{sec:bh-parameters}

We verify that all inequalities in Section~\ref{sec:bh-hardness} are
satisfied simultaneously by the explicit  assignment
\begin{align}
 \begin{gathered}
 M_0=256,\quad E=10^4,\quad U=10^{23},\\
 \sigma=10^{-9},\quad r=10^{-30},\quad
 \rho=10^{-60},\quad \zeta=10^{-5},\quad
 \eps_0=10^{-132},\quad \eta=\gamma=10^{-3}.
 \end{gathered}
 \label{eq:bh-parameters}
\end{align}
Notice that we have $M_0\geq 8G=648/5$,
$E\geq 8M_0+4=2052$, $U\geq 4E$, $r\leq\sigma/2$, and $\rho=r^2$. The amplification factor in~\eqref{eq:bh-source-decoding} is
\begin{align*}
 \cC(256)=\frac{164799}{400}<412,
\end{align*}
and the probability bound in~\eqref{eq:bh-menu-error} is
\begin{align*}
 \chi=\frac{6\cdot 10^4+64\cdot 10^{-7}+2\cdot 10^{-5}}{10^{14}}<10^{-9}.
\end{align*}
Consequently, the deviation error satisfies
\begin{align*}
 \Delta_{\mathrm{menu}}
 <10^{-5}+32\cdot 10^{-7}+2\cdot 10^{-5}+5\cdot 10^{-5}<10^{-4},
\end{align*}
giving maximum deviation gain in the original game at most
$412\cdot 10^{-4}+4\cdot 10^{-3}+3\cdot 10^{-3}=0.0482<1/20$. Moreover,
$\eps_0=10^{-132}\leq\min\cbr{\zeta/(11U),\rho^2/(100K^2)}$, and the
 scales in~\eqref{eq:bh-anchor-scales} are $C=10^{64}D$ and
$B=10^{128}C$.

\paragraph{Finite integer alphabet}
With $\eta=1/1000$, the external masses in~\eqref{eq:bh-decomposition} are integer multiples of $1/11000$. Since all menu prices are multiples of $1/15$ and the captive ratios $q/(1-q)$ have denominators dividing $12$, every unary payoff offset in~\eqref{eq:bh-intercepts} has denominator dividing $H_0 = 15 \cdot 11000 = 165000$, and the anchor attachment masses $W_0, W_1$ in~\eqref{eq:bh-menu-masses} have denominators dividing $12H_0$. Rescaling all market masses by the fixed integer $Q_0 = 144H_0 = 23{,}760{,}000$ leaves the normalized $\eps_0$-equilibrium condition unchanged while turning every positive market mass into an integer at most $5.94 \cdot 10^{223} < 10^{224}$.

\end{document}